\documentclass[journal=jacsat,manuscript=article]{achemso}
\SectionNumbersOn

\usepackage[version=3]{mhchem} 
\usepackage{xcolor}
\usepackage{bm}
\usepackage{amsfonts,amsmath,amssymb}
\usepackage{leftindex}
\usepackage{hyperref}
\hypersetup{
    colorlinks=true,
    citecolor=red,
    linkcolor=blue,
    urlcolor=cyan,
}

\newcommand{\opp}{\widehat{\bm{p}}}
\newcommand{\opa}{\widehat{\bm{a}}}
\newcommand{\opH}{\widehat{H}}

\newcommand{\opA}{\widehat{\bm{A}}}

\newcommand{\bmR}{\bm{R}}

\newcommand{\bmr}{\bm{r}}

\usepackage{tikz-feynman}
\usepackage{subcaption}

\author{Reza Karimpour}
\affiliation{Department of Physics and Materials Science, University of Luxembourg, L-1511 Luxembourg City, Luxembourg}
\author{Matteo Gori}
\affiliation{Department of Physics and Materials Science, University of Luxembourg, L-1511 Luxembourg City, Luxembourg}
\author{Alexandre Tkatchenko}
\affiliation{Department of Physics and Materials Science, University of Luxembourg, L-1511 Luxembourg City, Luxembourg}
\email{alexandre.tkatchenko@uni.lu}

\title{Quantum Field Approaches to Chemical Systems}

\begin{document}







\begin{abstract}


Quantum-matter theory (QMT), based on the Schrödinger or Dirac equations, is firmly established for both intra- and intermolecular interactions. However, there are two key issues with QMT. First, its applicability
to large molecular complexes is hindered by the relatively high computational cost of the calculations required to achieve high accuracy. Second, fields are also quantum objects that produce many intriguing effects beyond standard QMT approaches to molecular systems. This review focuses on recent developments in quantum-field theory (QFT) approaches to both covalent and non-covalent interactions for molecules in vacuum and subject to environments such as cavities and solvents. QFT provides a rich playground for novel chemical theories and insights. For example, chemical reactions and van der Waals interactions can be manipulated by cavities, boundaries, and optical excitations; novel interactions emerge when molecules interact with quantized fields; systems with millions of atoms could soon be treated with coarse-grained QFT formalisms; and unexpected scaling laws for atomic and molecular properties can emerge when QFT is applied to sets of chemical systems. This review sets the stage for an exciting QFT-driven path for further development of chemical theory.
\end{abstract}

\section{Introduction}
Our ability to model and understand chemistry is intimately tied to the development of quantum mechanics for interacting matter. Quantum-matter theory (QMT), based on the Schrödinger or Dirac equations, is the foundation of current computational methods and conceptual tools for modeling and understanding molecules and materials. The basis for QMT is the Hamiltonian of coupled nuclear and electronic degrees of freedom. Nuclear and electronic charges interact via the instantaneous Coulomb potential, giving rise to many-particle states, including molecular orbitals, molecular vibrations, and molecular plasmons (collective oscillations of the electron density). As many physicists and chemists know, this is an approximation~\cite{Cohen-Tannoudji1997, Milonni-QVac-1994, Lifshitz_QEDBook_1982, Craig1994, Salam2009, Buhmann-Disp-I-2013}. The quantum vacuum, as the ground state of interacting fluctuating fields of various types, is the fundamental object endowed with its own dynamical degrees of freedom. Quantum field theory (QFT) is the fundamental approach that provides tools for quantizing both matter and fields~\cite{Cohen-Tannoudji1997, Milonni-QVac-1994, Lifshitz_QEDBook_1982, PeskinQFT}. QFT approaches are not widely used in chemistry because treating the bound states in atoms and molecules is nontrivial in this theory. However, this is slowly changing in the 21st century, with many groups contributing to the brewing of a ``chemical QFT revolution''. The importance of QFT techniques is also highlighted by significant advances in quantum information and quantum computing applied to molecular systems~\cite{Aspuru-Science-2005, Aspuru-ChemRev-2019, Aspuru-RevModPhys-2020, MSarkis-PRResearch-2023}.

Within QMT, interactions in matter are typically separated into intra- and intermolecular components and are widely studied under different categories, defined by bond types and energy scales~\cite{Stone2013, Kaplan2006, Israelachvili2010}. However, the applicability of QMT to large molecular systems is limited by the immense computational cost of the calculations required to achieve the desired accuracy. Furthermore, molecules in cavities or subject to excitations (electric and magnetic fields as well as light) often require \textit{ad hoc} approaches beyond the static Schrödinger equation. Painstaking QM calculations based on coupled-cluster (CC) theory up to triple excitations and/or the quantum Monte Carlo (QMC) method can now reach an unprecedented accuracy of 1 kJ/mol (0.25 kcal/mol or 10 meV) per molecule for systems with a few dozen atoms or molecular crystals~\cite{ERC5, ERC6, ERC7}. For several important classes of molecular systems, much more efficient semilocal density-functional theory (DFT), including nonlocal many-body dispersion interactions, can achieve predictive accuracy comparable to that of CC or QMC methods when compared with experimental reference data~\cite{ERC8, ERC9}. However, DFT methods are routinely applicable to systems with only 100--1000 atoms.

Despite substantial progress in QMT method development, many fundamental questions remain about the completeness of QMT based on electronic Hamiltonians for large molecular systems~\cite{Venkataram-SciAdv2019-Nuclei-Vibration-vdW-Casimir, Venkataram-PRL2017-Unify-vdW-Casimir, Flick_PRL_variational_QED}, and several puzzles persist even when comparing apparently well-established benchmark QM calculations~\cite{ERC14}. For example, reference CC and QMC calculations are in mutual agreement on the binding energies of molecular complexes containing up to approximately 100 atoms; however, these ``benchmark'' methods begin to disagree when molecules form supramolecular complexes with a significant contribution from vdW dispersion interactions to their stability~\cite{ERC14}. For a cycloparaphenylene-ring/buckyball complex containing 132 atoms, the minimal disagreement between the state-of-the-art CC and QMC calculations reaches 31 kJ/mol~\cite{ERC14}. Identifying the reasons behind this unsettling discrepancy requires an improved understanding of electron correlations in many-particle systems over a wide range of spatial scales.

\begin{figure}[h!]
    \centering\hrulefill\par
    \begin{subfigure}[c]{\textwidth}
        \centering
        \includegraphics[width=0.88\textwidth]{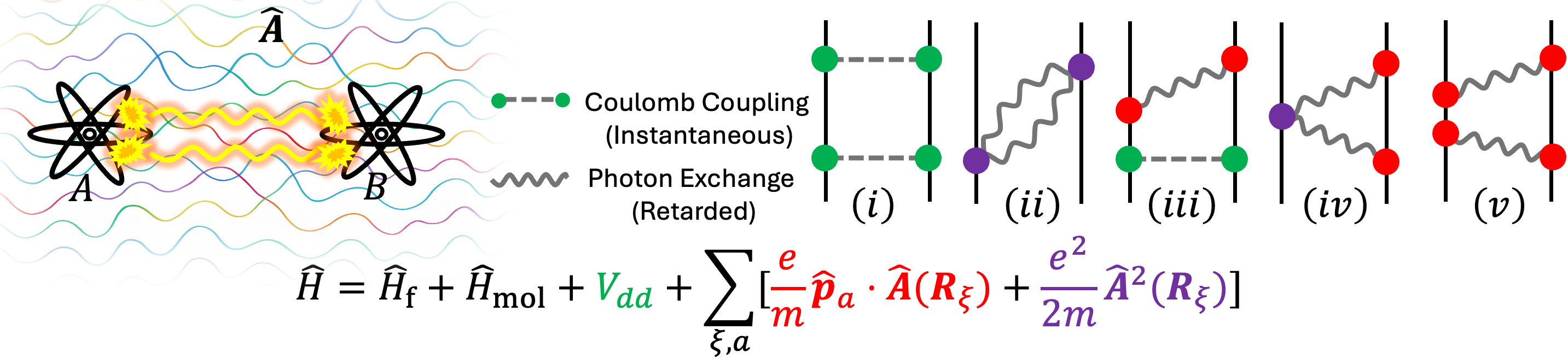}
        \vspace{-0.2cm}
        \caption{\vspace{-0.3cm}}
        \hrulefill\par
        \label{fig:dispersion-int}
    \end{subfigure}
    \begin{subfigure}[c]{\textwidth}
        \centering
        \includegraphics[width=0.9\textwidth]{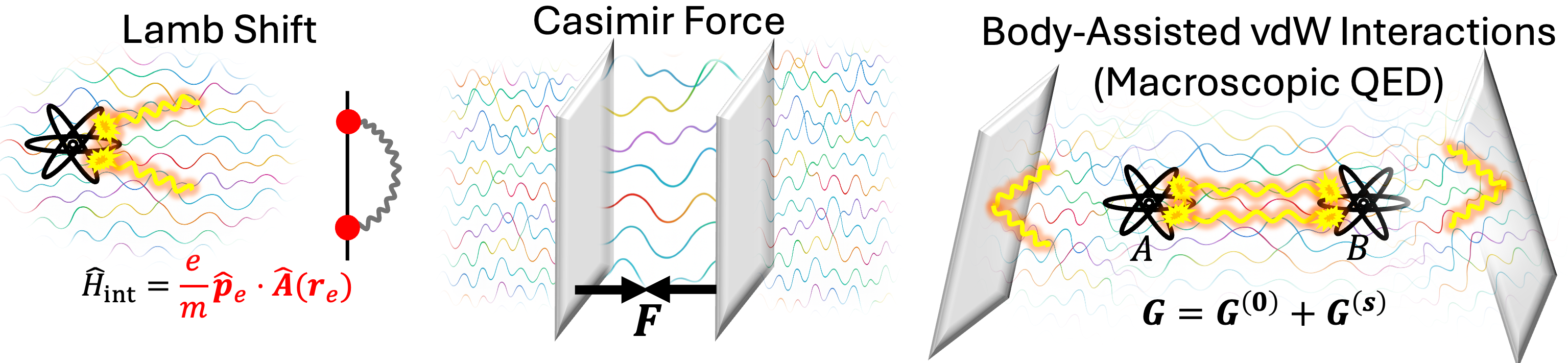}
        \vspace{-0.2cm}
        \caption{\vspace{-0.3cm}}
        \hrulefill\par
        \label{fig:QED-effects}
    \end{subfigure}
    \begin{subfigure}[t!]{0.4\textwidth}
        \centering
        \includegraphics[width=0.6\textwidth]{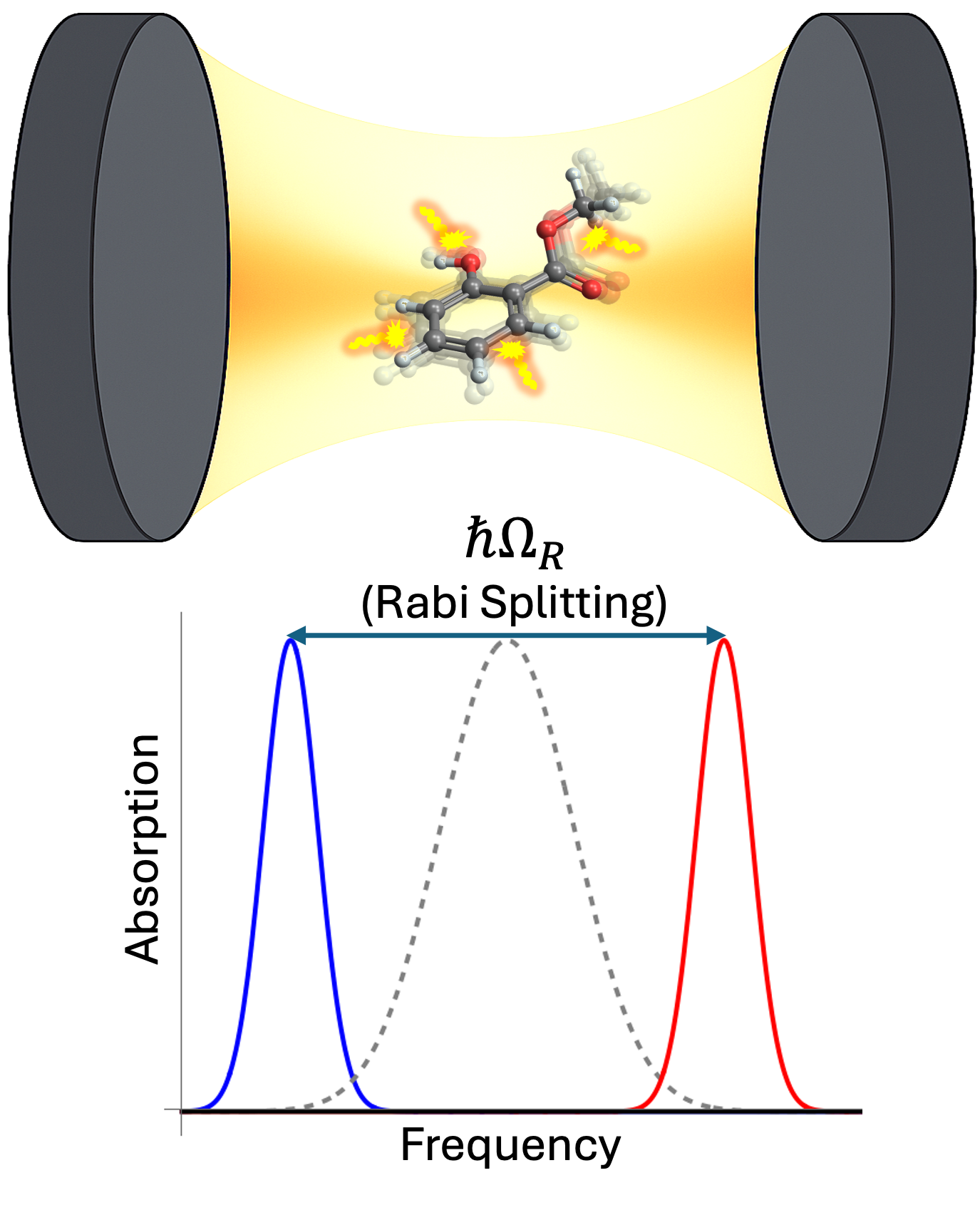}
        \caption{}
        \label{fig:polaritons}
    \end{subfigure}
    \vrule
    \begin{subfigure}[t!]{0.4\textwidth}
        \centering
        \includegraphics[width=0.6\textwidth]{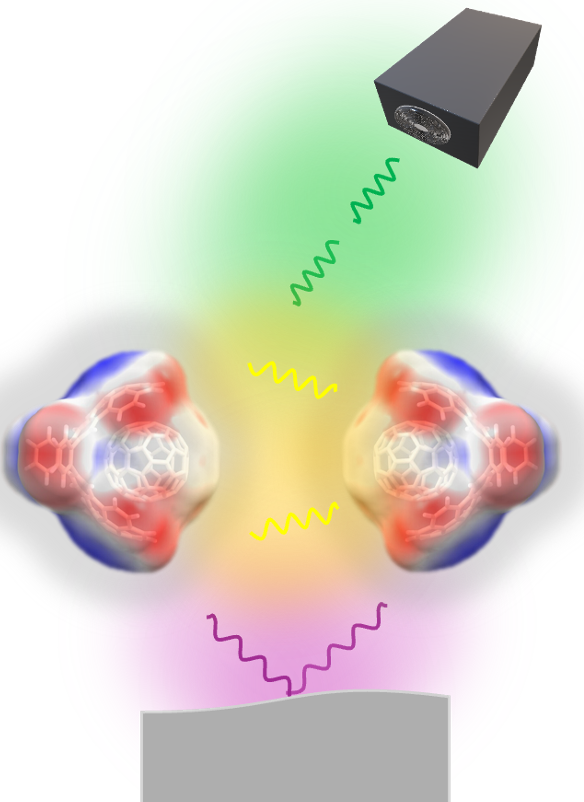}
        \caption{}
        \label{fig:QFT-Dream}
    \end{subfigure}\\
    \hrulefill\par
    \caption{\footnotesize \textbf{Molecular Interactions from a Quantum Field Theoretical Perspective.} 
    \textbf{(a)} Molecular interactions between two polarizable entities and their corresponding QED Feynman diagrams (up to the fourth-order perturbation theory) in the minimal-coupling formalism. Here, $\widehat{H}_{\rm f}$ is the Hamiltonian of the fluctuating electromagnetic field (EMF), $\widehat{H}_{\rm mol}$ denotes the atomic/molecular Hamiltonians, $V_{dd}$ is the dipolar Coulomb coupling, and the remaining sum describes atom–field interactions (see appendix~\ref{sec:Appendix_B} for more details about the Hamiltonian of a system of intercting matter and a quantized electromagnetic field). The diagrams represent processes up to fourth order in perturbation theory.
    \textbf{(b)} The well-known effects of the vacuum EMF on atoms/molecules, dispersive macroscopic bodies, or the combination of both. The presence of macroscopic bodies and boundaries alters EMF fluctuations, thereby influencing molecular interactions. In macroscopic QED, electric, magnetic, and geometric properties of the boundaries and macroscopic bodies or the bulk materials are encoded into classical Green functions, and the EMF is expressed in terms of that.
    \textbf{(c)} Inside optical cavities and resonators, strong matter-field couplings are reachable. In such situations, strong coupling between matter and the cavity EMF gives rise to hybrid states that combine properties of molecular and field states, known as polaritons. 
    \textbf{(d)} A quantum-field-theoretical description of matter as a field interacting with other fields (\textit{e.g.} EMF) allows for the full account of the many-body nature of molecular interactions under the influence of external fields and boundaries and enables the development of computational methods scalable from simple atomic dimers to complex biological macromolecules.}
\end{figure}

On a more fundamental level, interactions between particles are consequences of their coupling to the quantum fluctuations of vacuum fields. Such fluctuations occur at various scales, but the most relevant to atomic and molecular forces are vacuum electromagnetic field fluctuations~\cite{Cohen-Tannoudji1997, Milonni-QVac-1994, Salam2009, Buhmann-Disp-I-2013, Craig1994, Lifshitz_QEDBook_1982}. However, QMT treats atoms and molecules as closed quantum-mechanical systems and assumes that electromagnetic fields are fixed classical perturbations\cite{Salam2009}. Although this simplification makes the theory convenient, it also means that the field is not part of the quantum system. Therefore, this semi-classical theory cannot consistently describe essential features, such as vacuum fluctuations and emergent phenomena, including finite speed-of-light effects in dispersion interactions (see Fig.~\ref{fig:dispersion-int}), the Lamb shift (Fig.~\ref{fig:QED-effects}), spontaneous photon emission, the well-known Casimir effect between macroscopic ensembles of atoms and molecules (Fig.~\ref{fig:QED-effects}), and the intricate interplay between electronic and photonic degrees of freedom. These phenomena become increasingly pronounced in large systems, strongly coupled molecular ensembles, and in external electromagnetic fields or optical cavity environments~\cite{Salam2009, Thirunamachandran1980-external-field, Flick_natrev_2018, Flick_ACS_photonics_2019, Flick-NanoPhotonic2018-Strong-Rev, Rubio-JCP2021-interactions-in-cavities, flick2018ab, Rubio-2018SciAdv-cavity-QEDFT-superconductivity}. The limitations of semi-classical theory become particularly severe when dealing with modern coherent light sources and precision spectroscopy, where the quantized nature of the vacuum field assumes a central role.

Quantum electrodynamics (QED) overcomes these deficiencies by quantizing both matter and the radiation field within a unified dynamical framework. This methodology provides a rigorous foundation for electron–photon interactions, fully accounting for processes mediated by virtual and real photons. For instance, incorporating QED effects modifies fundamental interatomic dimer potentials, such as altering the vdW dispersion interaction scaling from $R^{-6}$ to $R^{-7}$ due to retardation~\cite{Casimir_Polder1948} at large distances $R$, or to $R^{-1}$ and $R^{-2}$ due to externally-induced excitations of the fluctuating electromagnetic field~\cite{Thirunamachandran1980-external-field}.

Importantly, QED has proven extraordinarily successful in practice. It not only offers conceptual completeness, but also delivers numerical predictions that agree with experiments across a vast range of scales, establishing it as the most accurate physical theory currently available. The QED methodology provides conceptual clarity, enabling a deeper understanding and predictive modeling of quantum correlations beyond traditional particle-like treatments. Recent advances, exemplified by methods such as quantum-electrodynamical density functional theory (QEDFT)~\cite{Rubio-PRA2014-QEDFT, Rubio-PNAS2015-Kohn-Sham-QEDFT, Rubio-PRA2018-bridging-Qchemistry-Qoptics, Rubio-QEDFT-Rev-2019, Rubio-JCP2021-interactions-in-cavities, Rubio2025-Dirac-Kohn-Sham} and coupled quantum Drude oscillator models (cQDO)~\cite{tkatchenko2012accurate, Fedorov2018, Vaccarelli2021, gori2023second, Loris-arxive2025, Almaz-JCP2025-Rev-QDO}, bridge the gap between highly accurate but computationally demanding techniques (e.g., coupled-cluster theory and quantum Monte Carlo methods) and more efficient but approximate density functional theory (DFT) approaches.

The inclusion of quantized electromagnetic fields and their interactions with matter fundamentally modify the electronic structure and response properties of molecules~\cite{Lamb-1947-Microwave-Method, Lamb1949-self-energy-bound-electron, Lamb1950-part1, Lamb1951-part2, FrenchWisskopf_LambShift1949, Drake1982_LambShift, Ebbesen2016-Perspective-Hybrid-Light-Matter, Huo2023-Review-PolaritonChemistry, Koch2024-Polaritonic-Response-Theory, Koch2024-Aromaticity, Feist2015-Mol-Structure-Bond-Length, Feist2017-Polaritonic-Chemistry-Organic-Molecules}. The significance of such alterations depends on the strength of the photon-matter coupling, which can range from weak to strong and ultrastrong. 
In the weak-coupling regime, the bare states of atoms and molecules provide a sufficiently accurate description of matter, and the quantized field acts as a perturbation that yields radiative corrections. Well-known examples of phenomena arising from weak atom-field couplings include spontaneous emission, the Lamb shift, Casimir-Polder interactions, and the electron's anomalous magnetic moment~\cite {Cohen-Tannoudji1997, Milonni-QVac-1994, Craig1994, Salam2009, Lifshitz_QEDBook_1982}. 
In contrast, strong and ultrastrong atom-field couplings lead to the formation of photon-matter hybrid states, known as polaritons~\cite{Ebbesen2016-Perspective-Hybrid-Light-Matter, Huo2023-Review-PolaritonChemistry, Feist2017-Polaritonic-Chemistry-Organic-Molecules, Koch2024-Polaritonic-Response-Theory}, which give rise to modified potential-energy surfaces and reshaped chemical landscapes. The ability to engineer these surfaces--for example, in optical cavities by tuning properties such as cavity frequency and coupling strength--is the central concept underpinning the emerging field of polariton chemistry~\cite{Ebbesen2016-Perspective-Hybrid-Light-Matter, Feist2017-Polaritonic-Chemistry-Organic-Molecules, Koch2024-Polaritonic-Response-Theory}.

An unambiguous signature of strong coupling is Rabi splitting in the absorption spectrum of materials inside cavities (as illustrated in Fig.~\ref{fig:polaritons}), where the material's excitation and the cavity photons are tuned into resonance, forming upper and lower polariton states~\cite{Thompson1992-Rabi-Splitting, Weisbuch-1992-semiconductor-Rabi-splitting, Weisbuch-1994-RoomTemp-Rabi-splitting, Lidzey-1998-Organic-Rabi-splitting, Liu2014-Polariton-Graphene}. 
The formation of polaritons has profound implications, offering novel pathways to influence chemical reactivity~\cite{Ebbesen2012-Photochemical-Reactivity, Ebbesen2016-GroundState-Reactivity, Ebbesen2019-GroundState-Reactivity, Ebbesen2019-Chemical-Reactivity, Fregoni2018-Azobenzene-Photoisomerization, Ahn2023-GS-Reactivity-Infrared} and conductivity~\cite{Orgiu2015-Conductivity, Bagliani2018-Conductivity, Rubio2022-Free-electron-gas-conductivity, Appugliese2022-conductivity}, shifting conical intersections~\cite{Vendrell2018-Conical-Intersection, Vibok2018-Conical-Intersection, Gu2020-Conical-Intersection}, or modifying molecular bond lengths\cite{Feist2015-Mol-Structure-Bond-Length, Schnappinger2023-Bond-Length, Schnappinger2024-Bond-Length}. Beyond coupling photonic states to molecular electronic or vibrational states, the electromagnetic nature of the fluctuating field also enables matter-field interactions via the magnetic degrees of freedom of these systems. Coupling the quantized field to molecular spins, which are the engines of their magnetic properties, alters spin interactions ({\it e.g.} via field-mediated spin-orbit coupling) and effectively retunes the molecules' magnetic responses, thereby directly influencing their aromaticity~\cite{Koch2024-Polaritonic-Response-Theory, Koch2024-Aromaticity, Deng2014-Spin-Orbit}. 
A natural framework accounting for such interactions is relativistic quantum electrodynamics, in which spin is inherently incorporated into the quantum-mechanical description of matter through the Dirac equation or its multiple-fermion extensions~\cite{Koch2025-Relativistic-QED}. Another advantage of employing relativistic QED is its ability to achieve precision in describing atomic and molecular systems at the level of fundamental constants, where relativistic and radiative effects become inseparable and measurable in the domain of precision spectroscopy~\cite{Matyus2024-Perspective-Precision-Physics, Matyus2024-rQED-Correction-1photon, Matyus2023-Pre-Born-Openheimer-leading, Matyus2023-Pre-Born-Openheimer-2Body}.

In addition to external electromagnetic fields, collective interactions among neighboring atoms and molecules strongly influence the molecular response properties. As systems grow in size, many-body interactions--particularly van der Waals dispersion forces--modify the local electronic environment and can significantly enhance molecular polarizabilities. This enhancement is a genuine many-body effect: in one-dimensional carbyne chains, for example, the polarizability of each carbon atom increases by a factor of 50 along the extended structure due to cooperative dispersion interactions~\cite{Gobre-NatComm2013, ambrosetti2016wavelike}. Similar trends are observed in other extended materials, such as graphene, where atoms near the interior of a sheet exhibit larger in-plane polarizabilities than those at the edges~\cite{Gobre-NatComm2013}. Consequently, atoms or molecules embedded in large, correlated matter systems interact more strongly with the fluctuating electromagnetic field than their isolated counterparts, thereby amplifying QED effects and radiative corrections. Any realistic treatment of light–matter coupling in extended systems must therefore incorporate these many-body enhancements in the matter subsystem. 

In quantum chemistry, approaches that combine density-functional theory with many-body dispersion (MBD) methods~\cite{tkatchenko2012accurate, DiStasio-2014-MBD-Review, Reilly2015-MBD, gori2023second} successfully capture these cooperative effects for large molecular assemblies. However, when the electromagnetic field itself is quantized, such techniques no longer apply directly. From a QFT standpoint, matter is naturally described as a field whose excitations constitute the molecular structure, so many-body correlations arise intrinsically within the formalism. A comprehensive description of molecules interacting with quantized electromagnetic fields--whether in cavity vacuum conditions or under external driving--thus becomes a problem of interacting quantum fields (see fig.~\ref{fig:QFT-Dream} for an illustration of such a \textit{chemical QFT} end goal). The advantage of this viewpoint is that the theoretical framework remains unified and scalable, regardless of whether one studies a small organic molecule or a complex macromolecular environment. As experimental control of large molecular systems with confined or structured light continues to advance, the need for a QFT-grounded, many-body–consistent description of molecular QED becomes increasingly apparent.



In the following, we first summarize the prominent experimental evidence demonstrating the impact of the quantum vacuum on chemical systems. We then present a perspective on QED and QFT methodologies in quantum chemistry. We begin by discussing the conventional quantum chemistry approach for describing the electronic structure of molecular systems. Subsequently, we explore the advantages of employing field-theoretical methods, such as second quantization and the introduction of particle density (as a scalar field) and polarization tensor fields, in quantum chemistry. We then delve into the description of atoms and molecules in molecular QED, highlighting how QED effects influence non-covalent and covalent interactions (e.g., rotational and vibrational properties of bonds). Finally, we emphasize the importance of QED effects in precision experiments and the key considerations for developing predictive theories for both strong and weak QED effects.

\section{Concise Summary of Experimental Evidence for Quantum Field Effects in Chemical Systems}
From a historical perspective, the most remarkable evidence that a comprehensive theory was required beyond QMT (Schrödinger or Dirac equation for matter) emerged with the observation of splitting in the $2S_{1/2} - 2P_{1/2}$ energy levels of the hydrogen atom~\cite{Milonni-QVac-1994, Craig1994, Lamb1950-part1, Lamb1951-part2, Lamb-1947-Microwave-Method, Lamb1949-self-energy-bound-electron, Drake1982_LambShift, FrenchWisskopf_LambShift1949}. The QMT, even when employing the Dirac equation, predicts degenerate $2S_{1/2}$ and $2P_{1/2}$ levels for hydrogen~\cite{Sakurai_advancedQM_1967}. Lamb and Retherford, using microwave spectroscopy~\cite{Lamb-1947-Microwave-Method}, measured a small energy difference between these states, with the $2S_{1/2}$ state having a slightly higher energy than the $2P_{1/2}$ state~\cite{Lamb1950-part1, Lamb1951-part2, Lamb1949-self-energy-bound-electron}. The inability of the QMT to account for the Lamb shift arises from its neglect of the interaction between the bound electron and the EMF. One interpretation of the Lamb shift attributes it to the atom's coupling to the quantized EMF fluctuations, as illustrated in Fig.~\ref{fig:QED-effects}, which induce variations in the electron's position and thereby modify the potential it experiences~\cite{Milonni-QVac-1994}. Experimental measurements demonstrated that these fluctuations produce energy shifts that are most pronounced in S states, splitting degenerate levels, with the largest observed shift of 1057.8298 MHz for the $2S_{1/2}-2P_{1/2}$ transition (roughly 10$^{-4}$ kcal/mol). Although this shift appears insignificant in a hydrogen atom, it is worth mentioning that a muonic hydrogen exhibits a Lamb shift of approximately 4.8 kcal/mol due to the significantly higher mass of the muon~\cite{Muonic-Lamb-shift-2007}. The scaling of the Lamb shift for many-body systems beyond atoms remains unknown at present. 

Beyond EMF, the uncertainty principle in quantum mechanics permits fluctuations in all other quantum fields, including both bosonic and fermionic ones. Within quantum field theory, massive fermionic fields are a natural generalization of the massless electromagnetic field. While photons can be created and annihilated in quantum electrodynamics (QED), general field theories also allow for the creation and annihilation of other massive or massless particles, such as Dirac fermions~\cite{Cohen-Tannoudji1997, Milonni-QVac-1994, PeskinQFT, Lifshitz_QEDBook_1982}. Early evidence for fermionic vacuum effects was provided by Heisenberg and Euler, who examined the low-energy effective action of QED in a constant electromagnetic background and demonstrated that virtual electron-positron pairs polarize the vacuum~\cite{Milonni-QVac-1994}. Subsequent calculations and measurements of the Lamb shift~\cite{Lamb1949-self-energy-bound-electron, Drake1982_LambShift} confirmed that vacuum polarization induces spectral shifts, such as the $2S_{\frac{1}{2}}-2P_{\frac{1}{2}}$ splitting of approximately 27~MHz in hydrogen. Although this represents only a fraction of the total Lamb shift in an ordinary hydrogen atom, it is crucial to ensure that theory and experiment match. These effects are particularly significant in muonic atoms, where vacuum polarization dominates, leading to a substantial increase in the vacuum-polarization contribution to the Lamb shift. For example, in muonic helium, vacuum polarization accounts for approximately $90\%$ of the Lamb shift~\cite{Glauber1960-Lamb-shift-vacuum-polarization}.

Another notable early achievement of QED was its explanation of spontaneous emission. Semiclassical theory suggests that an atom not subjected to an external electromagnetic field does not radiate. Nevertheless, it is well established that a free atom in an excited state will eventually decay to its ground state by emitting a photon. In QED, where the vacuum is characterized by a fluctuating electromagnetic field, this emission is not entirely spontaneous, but is instead stimulated by the vacuum field~\cite{Craig1994, Milonni-QVac-1994}. The recognition that spontaneous emission can be significantly enhanced or suppressed by the surrounding environment, as proposed by Purcell and termed the Purcell effect, inspired the concept of engineering the vacuum to control matter's behavior. This insight ultimately led to the development of Cavity-QED~\cite{Haroche2006-exploring-the-quantum}, a field that has become prominent among physicists and chemists for investigating both weak and strong matter-field interactions.

Beyond isolated atoms, precision spectroscopy of small molecules provides a rigorous testbed for QED. Recent measurements of the hydrogen molecule and molecular helium ion ($^4He_2^+$) have reached unprecedented accuracy, challenging theoretical models. For example, the lowest-energy rotational interval of $^4He_2^+$ has been measured as $70.937589\,\text{cm}^{-1}$, establishing a benchmark for three-body QED calculations~\cite{Matyus-PRL2020, Matyus2024-Perspective-Precision-Physics}. Discrepancies between theoretical predictions and experimental results in these systems often indicate the need for higher-order QED corrections. In the HD molecule, precision measurements have revealed deviations of $1.4\sigma$ to $1.9\sigma$ relative to current theoretical values, underscoring the complexity of non-adiabatic and relativistic QED effects in chemical bonds~\cite{Silkowski2023-QED-Effects-H2}. Furthermore, the \textit{proton radius puzzle}, a $4\%$ discrepancy in the proton charge radius measured via electronic versus muonic hydrogen spectroscopy, highlights the sensitivity of chemical systems to vacuum field interactions~\cite{Matyus2023-Pre-Born-Openheimer-2Body}. These findings confirm that even seemingly simple molecular properties are fundamentally connected to fluctuations in the underlying quantum fields~\cite{Matyus2024-Perspective-Precision-Physics}.

In complex chemical systems (more intricate than small molecules or single atoms) confined within cavities, as shown in Fig.~\ref{fig:polaritons}, strong coupling between molecules and quantized fields has provided compelling experimental evidence that QED effects alter macroscopic chemical properties. The most direct indicator of strong coupling is the Rabi splitting of the molecular absorption spectrum into upper and lower polariton branches ($P_+$ and $P_-$), separated by the Rabi frequency $\hbar\Omega_R$. A prominent example of Rabi splitting was observed with cyanine dye molecules and J-aggregates (TDBC) in Fabry–Pérot cavities. In these systems, substantial Rabi splittings of up to approximately 1 eV, or about 0.25 of the resonance frequency, have been measured at room temperature~\cite{Ebbesen-ChemPhysChem2013-Polariton-Dynamics, Ebbesen2016-Perspective-Hybrid-Light-Matter, Feist2017-Polaritonic-Chemistry-Organic-Molecules} (see fig. \ref{fig:strong-hybrid}). These significant Rabi splittings, resulting from strong coupling of molecular electronic states to a cavity's vacuum field, have been shown to influence photochemical reactivity. For example, the photoisomerization reaction between spiropyran and mecrocyanine was modified by resonant coupling to a Fabry-Pérot cavity, resulting in a Rabi splitting of $\hbar\Omega_R = 700~ {\rm meV}$, which could create a polaritonic barrier that effectively inhibits the reaction rate~\cite{Ebbesen2012-Photochemical-Reactivity, Huo2023-Review-PolaritonChemistry}. 
Other studies have used plasmonic nanoantennas to generate a strong cavity field that couples to dye molecules, inducing large Rabi splitting that allows excited-state populations to decay faster than they would undergo triplet-state reactions, thereby suppressing the photobleaching of organic chromophores\cite{Munkhbat2018-SciAdv-suppress-photobleaching, Huo2023-Review-PolaritonChemistry}.

In addition to electronic transitions, strong coupling of molecules to cavity vacuum fields can also occur via their vibrational degrees of freedom, a phenomenon termed Vibrational Strong Coupling (VSC)~\cite{Ebbesen2016-Perspective-Hybrid-Light-Matter, Huo2023-Review-PolaritonChemistry}. By tuning a cavity to resonate with specific bond vibrations, VSC hybridizes molecular vibrational modes with the vacuum field, fundamentally restructuring the ground-state potential energy surface and thereby modifying chemical kinetics and product selectivity without external optical pumping. A fundamental example of this ``chemistry in the dark'' is the deprotection of 1-phenyl-2-trimethylsilylacetylene (PTA), where resonant coupling to its Si-C stretching mode at $860 {\rm cm}^{-1}$ produced a Rabi splitting of $98 {\rm cm}^{-1}$ and reduced the reaction rate by a factor of 5.5~\cite{Ebbesen2016-Perspective-Hybrid-Light-Matter, Huo2023-Review-PolaritonChemistry}. Such modifications, which may also include rate enhancement through solvent coupling or the realization of mode-selective chemistry~\cite{Huo2023-Review-PolaritonChemistry}, are typically enabled by collective coupling that scales with $\sqrt{N}$ for $N$ molecules~\cite{Huo2023-Review-PolaritonChemistry, Flick2022-chemical-reactivity}. These effects are frequently attributed to \textit{dynamical caging} by the cavity field and have recently been extended to the relativistic regime to control formally forbidden singlet-triplet transitions in heavy atoms such as mercury, providing a non-invasive approach to manipulate intersystem crossing and phosphorescence~\cite{Koch2025-Relativistic-QED, Rubio2025-Dirac-Kohn-Sham}.


It is important to stress that while the aforementioned experimental evidence focuses on specific atomic or molecular benchmarks, QED effects are fundamental and universal; they apply to all atoms, molecules, and materials characterized by many interacting electrons and nuclei. However, the precise scaling of QED phenomena for systems of increasing size remains an open question. In many-body systems, simple dipole-coupling models often fail to capture the non-additive nature of the interaction, much as the vdW dispersion energy deviates from the standard pairwise interatomic $R^{-6}$ power law in extended nanostructures~\cite{Gobre-NatComm2013}. From a QED perspective, rough scaling arguments based on the $\widehat{\bm{p}}\cdot\widehat{\bm{A}}$ interaction term suggest that the coupling strength is mediated by the square of the transition dipole moment ($|\bm{\mu}|^2$). Hence, in large polarizable systems, the delocalization of electrons and the emergence of collective modes can lead to a significant enhancement of this effective coupling, potentially making QED effects far more dominant than in isolated atoms. Identifying and extracting these contributions from experimental measurements remains a formidable challenge, as we currently lack a practical many-body theory capable of disentangling pure vacuum effects from intricate correlations that are omnipresent in chemical systems.

\section{Molecules And Fields}

The standard formulation of quantum chemistry 
models electrons as point-like quantum particles 
and nuclei as classical charges, interacting via the electrostatic 
Coulomb potential. While wavefunction-based methods built 
on this picture have been highly successful, the particle-oriented perspective exhibits intrinsic limitations, 
particularly in its ability to provide a unified and 
scalable framework spanning systems from small molecules 
to large biomolecular complexes in realistic environments. 
Subsection \ref{subsec:advlim_particle} reviews the 
strengths and limitations of particle-based descriptions 
of electronic degrees of freedom. Subsection 
\ref{subsec:QFT_intro} introduces a field-theoretical 
formulation in which both matter and interactions are 
described by the Schrödinger field for many-body fermionic 
systems and the quantized electromagnetic field. Finally, 
subsection \ref{subsec:nonrel_QED} presents the conceptual 
foundations of a more fundamental and unifying framework 
for non-relativistic quantum electrodynamics (meaning that 
matter degrees of freedom are treated non 
relativistically) in which electronic matter
is described by fields and interactions are mediated by
fully quantum dynamical degrees of freedom of the 
electromagnetic field.

\subsection{Advantages and Limitations of Particle-Based Matter Hamiltonians}
\label{subsec:advlim_particle}
The quantum-mechanical description of molecules and 
materials is commonly based on a particle-based picture in 
which electrons and atomic nuclei interact via the 
instantaneous electrostatic Coulomb potential. The 
significant mass disparity between nuclei and electrons 
leads to the usual separation of mass scales, making the 
clamped-nuclei approximation highly effective. Given this 
approach, nuclei---even when identical---are treated as 
fixed, distinguishable classical point charges, an 
assumption typically justified within the Born--
Oppenheimer approximation. 

In the first-quantized representation, the positions and 
momenta $\{\bm{r},\bm{p}\}$ of each particle are promoted 
to Hermitian operators $\{\hat{\bm{r}},\hat{\bm{p}}\}$ 
acting on the Hilbert space of square-integrable functions 
over real space. These operators satisfy the canonical 
commutation relations
$[\hat{r}_{k,x}, \hat{p}_{k',x'}] = i\hbar\, \delta_{kk'} 
\delta_{xx'} \hat{I}
$
and
$
[\hat{r}_{k,x}, \hat{r}_{k',x'}] = [\hat{p}_{k,x}, \hat{p}_{k',x'}] = 0 $
where $k, k'$ label particles and $x, x'$ label Cartesian coordinates.
Neglecting the trivial Coulomb repulsion 
between nuclei, the molecular 
Hamiltonian takes the form
\begin{equation}
\label{eq:Coulel_Ham}
\hat H_{\rm el-Coul}=
\sum_{i=1}^{N_e}\underbrace{\frac{\|\hat{\bm p}_i\|^2}{2m_e}}_{\widehat K_{\rm el}[\hat{\bm{p}}_i]}
+\dfrac{1}{2}\sum_{\substack{i,j=1\\ i\neq j}}^{N_e}\underbrace{\frac{e^2}{4\pi\varepsilon_0\|\hat{\bm r}_i-\hat{\bm r}_j\|}}_{\widehat U^{\rm coul}_{\rm el-el}[\hat{\bm{r}}_i,\hat{\bm{r}}_j]}
-\sum_{i=1}^{N_e}\sum_{A=1}^{N_n}\underbrace{\frac{Z_Ae^2}{4\pi\varepsilon_0\|\hat{\bm r}_i-\bm R_A\|}}_{\widehat U^{\rm coul}_{\rm n-el}[\hat{\bm{r}}_i,\bm{R}_A]}.
\end{equation}
Here, $\widehat K_{\rm el}$ denotes the single-particle kinetic energy operator, $\widehat U^{\rm coul}_{\rm el-el}$ the electron--electron Coulomb interaction, and $\widehat U^{\rm coul}_{\rm n-el}$ the electron--nuclear interaction potential, where $\bm R_A\in\mathbb R^3$ and $Z_A$ denotes the position and atomic number of the $A$-th nucleus. Equivalent expressions for the nucleus--electron and electron--electron interactions are
\begin{align}
\label{eq:Coulel_pot}
&\widehat U^{\rm coul}_{\rm n-el}=
\iint\hat\rho_{\rm el}(\bm r)\rho_{\rm n}(\bm r')
V_{\rm Coul}(\bm r,\bm r')\,\mathrm d^3\bm r\,\mathrm d^3\bm r'= \\
&\widehat U^{\rm coul}_{\rm el-el}=\dfrac{1}{2}
\iint :\hat\rho_{\rm el}(\bm r)\hat\rho_{\rm el}(\bm r'):
V_{\rm Coul}(\bm r,\bm r')\,\mathrm d^3\bm r\,\mathrm d^3\bm r', 
\end{align}
where $V_{\rm Coul}(\bm r,\bm r')=(4\pi\varepsilon_0\|\bm r-\bm r'\|)^{-1}$ is the kernel modeling electrostatic Coulomb interactions, $\hat\rho_{\rm el}(\bm r)=-e\sum_{a=1}^{N_e}\delta^3(\hat{\bm r}_a-\bm r)$ is the electronic density operator, and $\rho_{\rm n}(\bm r)=e\sum_{A=1}^{N_n}Z_A\delta^3(\bm R_A-\bm r)$ is the nuclear density. Note that self-interaction terms are unphysical and must be removed via regularization of the electronic density product, i.e. $:\hat{\rho}_{\rm el}(\bm{r})\hat{\rho}_{\rm el}(\bm{r}'):=\hat{\rho}_{\rm el}(\bm{r})\hat{\rho}_{\rm el}(\bm{r}')-(-e)\hat{\rho}_{\rm el}(\bm{r}) \delta^3(\bm{r}-\bm{r}') $. 
It should be noted that all Coulomb interactions in the system can 
be rewritten in terms of the interaction between the electronic charge
density field $\hat\rho_{\rm el}(\bm r)=-e\sum_{a=1}^{N_e}\delta^3(\hat{\bm r}_a-\bm r)$ and the electrostatic potential $\hat{V}(\bm{r},t)$
\begin{equation}
\widehat U^{\rm coul} = \widehat U^{\rm coul}_{\rm n-el} + \widehat U^{\rm coul}_{\rm el-el} =\dfrac{1}{2} \int \hat{\rho}_{\rm{Tot}}(\bm{r}) \hat{V}(\bm{r}) \,\mathrm d^3\bm r -  \dfrac{1}{2}
\iint\rho_{\rm n}(\bm r)\rho_{\rm n}(\bm r')
V_{\rm Coul}(\bm r,\bm r')\,\mathrm d^3\bm r\,\mathrm d^3\bm r'
\end{equation}
where $\hat{V}[\hat{\rho}_{\rm Tot}(\bm{r})]$ is a functional of the 
total charge density operator $\hat{\rho}_{\rm Tot}(\bm{r})=\hat\rho_{\rm el}(\bm r)+\rho_{\rm n}(\bm r)$ that satisfies the Poisson equations, i.e.
\begin{equation}
    \nabla^2 \hat{V}(\bm{r}) = -\dfrac{\hat{\rho}_{\rm Tot}(\bm{r})}{\varepsilon_0}\,\Longrightarrow \hat{V}(\bm{r}) = \int V_{\rm Coul}(\bm r,\bm r') \hat{\rho}_{\rm Tot}(\bm{r}') \,\,\mathrm{d}^3\bm{r}' .
\end{equation}
Such a formulation of the Coulomb Hamiltonian suggests that at a more 
fundamental level the interactions between matter 
degrees of freedom should be described as interactions 
between \textit{fields}, e.g. the charge density field 
and the electromagnetic field (EMF).

Electrons are spin-$\tfrac12$ fermions, and the associated electronic wavefunction $\psi(\{\bm{r}_a,s_a\}_{a=1,\ldots,N_e})$ is defined as the projection of the quantum state 
onto position--spin eigenstates. A wide range of seminal computational approaches  \cite{cramer2013essentials,tarczay2001anatomy,perdew2001jacob,
akimov2015large,austin2012quantum,chan2011density,eriksen2020ground} have been developed to solve the time-independent Schr\"odinger equation associated with Eq.~\eqref{eq:Coulel_Ham}, which constitutes the foundational equation of quantum chemistry in first quantization. 

Recent advances in computational 
methods and experiments have 
highlighted limitations of this 
particle-based description, which 
can be grouped into two 
categories: (i) the representation 
of electronic degrees of freedom, 
and (ii) the treatment of the 
Coulomb interaction as a static, non-dynamical field. 

In first quantization, spin statistics are not encoded in the 
observable algebra and must 
instead be imposed by 
antisymmetrizing the electronic 
wavefunction. As a result, the 
many-electron wavefunction is 
defined on a $4N_e$-dimensional 
configuration space, severely 
constraining the construction of 
systematically improvable and 
scalable ans\"atze\cite{szabo1996modern,helgaker2000molecular}.
This leads to exponential scaling in wavefunction-based methods like Full Configuration Interaction, as well as expressivity limitations in Variational Quantum Monte Carlo~\cite{becca2017quantum}. These challenges are partially alleviated by the growing functional expressivity offered by the neural-network wavefunction ans\"atze~\cite{pfau2020ab,hermann2023ab,gao2024distributed}.
Further complications arise when 
the electron number is not fixed, for 
instance, when exploring molecular
chemical space. In this case, the 
wavefunction must then be defined 
over configuration spaces of 
varying dimensionality.
Therefore, projection  procedures are required 
to evaluate real-space observables,
including the electron density, 
whose topological properties 
encode essential information about 
chemical bonding (see \cite{martinpendasTopologicalApproachesChemical2023} and references therein). 
In addition, particle-index-based 
partitions implicit in first-quantized formulations can lead 
to ambiguities when applying quantum-information measures to 
systems of identical fermions 
\cite{aliverti2024can} (see Section \ref{subsec:QFM_chembond} for further details).
Moreover, first-quantization picture is not suited 
for describing relativistic effects for electrons 
in atoms and molecules. 
Such relativistic effects have been shown to be 
particularly relevant for core properties of heavy 
atoms. For instance, relativistic effects on the 
ionization energy and electron affinity of the gold 
atom are comparable in magnitude to electron 
correlation effects, as estimated at the coupled-
cluster CCSD(T) 
level~\cite{pavsteka2017relativistic}. 
A consistent description of these effects, and the 
development of effective relativistic quantum 
Hamiltonians, require field-based approaches for 
matter~\cite{liu2020essentials}. 
These limitations motivate alternative 
representations of many-body quantum systems of 
matter particles in terms of fields defined in real 
or momentum space.

At the same time, the Hamiltonian in 
Eq.~\eqref{eq:Coulel_Ham} is structurally 
limited by its exclusive reliance on 
electrostatic Coulomb interactions. Here, the EMF does not appear as 
an independent dynamical quantum object, and energy/momentum exchange via its quantized 
excitations is excluded at the 
Hamiltonian level. 
These limitations are evident 
when matter degrees of freedom 
interact with real EMF excitations such as photons and/or electric and magnetic fields. This occurs, for example, with energy transfer in photosynthetic molecular complexes\cite{Craig1994,
craig1998chiral,salam2018unified,
franz2023macroscopic,babcock2024ultraviolet,patwa2024quantum}. 
However, EMF excitations are also relevant even without external photons. Intrinsic quantum EMF excitations are carried by ``virtual'' photons, which can affect both intramolecular and intermolecular interactions~\cite{Craig1994,Salam2009}.
Examples include 
large-scale charge displacements over long 
distances~\cite{ambrosetti2016wavelike,khabibrakhmanov2025noncovalent} in supramolecules or 
biomolecular complexes,
or when EMF–matter coupling
is enhanced by 
specific setups (e.g. molecules in optical 
cavities)\cite{flick2018ab,Huo2023-Review-PolaritonChemistry,hsu2025chemistry}, or when high 
computational accuracy is required to predict 
electronic, optical, or vibrational molecular spectra~\cite{Matyus2024-Perspective-Precision-Physics}.
Moreover, leading-order quantum electrodynamic effects for 
relativistic bound electrons—most notably the Lamb shift—have 
been conjectured, on the basis of rough estimates, to contribute 
on the order of $\sim 1$ kcal/mol to the energetics of small 
molecules containing heavy elements~\cite{dyall2001lamb}. The 
same effects have been shown to produce measurable
corrections to predicted single and double ionization energies of 
1s and 2s orbitals for elements from the third row of the 
periodic table~\cite{niskanen2017qed}.
Taken together, these considerations 
motivate a reformulation in which 
both matter and electromagnetic degrees of freedom are 
described within a QFT framework.
In the next section, we introduce the 
formalism of quantum fields, with 
particular emphasis on the 
Hamiltonian formulation of the 
electromagnetic field; the physical 
consequences and applications 
of this extended description 
are discussed in 
Secs.~\ref{sec:QFM_section} and \ref{sec:atmol_inQED}.


\subsection{Quantum Description of Matter and the Electromagnetic Field}
\label{subsec:QFT_intro}
Although often associated with high-energy or condensed-matter physics, 
quantum field theory (QFT) provides a universal language for describing 
a wide range of many-body quantum systems. It naturally encodes 
identical particles in 3D 
space and treats fields (like 
electromagnetism) as dynamical 
quantum entities. 
Crucially, this 
framework extends to quantum chemistry, 
offering new insights by unifying many-body 
effects and correlations under a unified formalism.~\cite{blaizot1986quantum,fetter2012quantum,altland2010condensed}

Conceptually, the particle-based 
first-quantized 
formalism is analogous to a 
Lagrangian description in fluid 
mechanics, keeping track of 
individual constituents explicitly, 
whereas QFT plays a role 
analogous to an Eulerian description, 
formulated in terms of (density) fields defined over space.

In a field-theoretic formulation, quantum 
degrees of freedom are described by fields 
defined over space.
Depending on their nature, such fields may 
transform as scalars, vectors, or spinors, and 
their dynamics is governed by partial 
differential equations in real space\cite{umezawa1982thermo,umezawa1995advanced,araki1999mathematical,blasone2011quantum}.
For practical purposes, a basis 
set of square-integrable field modes $\{\bm{\phi}_k(\bm{r})\}_{k=1,...+\infty}$ is introduced, 
typically chosen as eigenfunctions of a linear operator describing the corresponding free 
field, allowing a generic field expansion, i.e.
$\bm{\Phi}(\bm{r},t) = \sum_k c_k(t) \bm{\phi}_k(\bm{r})$.
Quantization is achieved by promoting the 
classical field amplitudes $c_k$ to operators 
acting on a Fock space, introducing annihilation 
and creation operators $\{\hat c_k, \hat c_k^\dagger\}_k$ and a vacuum 
state $|0\rangle$ defined by $\hat c_k |0\rangle = 0$. Acting on the vacuum, $\hat c_k^\dagger$ 
creates a single quantum of excitation in mode $k$. 
The states of the field are then 
represented in the
occupation-number basis $|n_1,n_2,\ldots,n_k,\ldots\rangle$, defined as eigenstates of the number operators $\hat n_k = \hat c_k^\dagger \hat c_k$. In QFT, the 
Heisenberg picture is often preferred: fields 
are treated as dynamical variables, while 
asymptotic quantum states (in Fock space) are 
time-independent. Although the interaction 
picture is commonly used for practical 
computations in chemical applications (e.g., 
perturbative response and Fermi’s golden rule), 
care is needed. Unlike quantum mechanics, 
the vacuum states of free and interacting 
Hamiltonians are generally unitarily inequivalent in the continuum limit — a 
consequence of Haag’s theorem\cite{haag1964algebraic,blasone2011quantum}. This formal 
subtlety underlines why QFT requires careful 
handling of interactions, 
even in non-relativistic chemical settings.
Spin--statistics are encoded directly in the operator 
algebra: bosonic fields satisfy canonical commutation relations,
while fermionic fields obey canonical anticommutation relations. This 
construction is referred to as \textit{second 
quantization}\cite{berazin2012method} and forms 
the basis of field-theoretic approaches to 
quantum many-body systems.

A key structural distinction between QFT and first-quantized quantum mechanics is particularly relevant for 
quantum-chemical applications across multiple length 
scales. In first quantization, spatial coordinates are 
promoted to operators, and no c-number variables directly 
represent real-space configurations.
In QFT, by contrast, field amplitudes are quantized 
while spatial coordinates remain c-numbers, enabling a direct
real-space description of structure formation and collective behavior, 
even for systems with arbitrary and fluctuating particle 
numbers~\cite{umezawa1995advanced}.
Correspondingly, field operators are labelled by spatial 
coordinates or mode indices rather than by particle indices. 
As a result, observables such as densities and currents 
admit a direct real-space representation, without the need 
for projection procedures. Within this formalism, single-
particle orbitals emerge as a choice of basis for expanding 
the field operators, rather than as explicit building blocks 
of the many-body wavefunction.
The specific field used to represent the matter degrees of freedom depends on the physical regime: relativistic electrons are described by the Dirac field, whereas non-relativistic electrons are described by the Schrödinger field (a concise summary of the principal conceptual aspects of quantum field theory is provided in Appendix~\ref{sec:Appendix_A}). As this review focuses primarily on a non-relativistic treatment of matter degrees of freedom, selected applications of non-relativistic field-based approaches to quantum chemistry are reviewed in Sec.~\ref{sec:QFM_section}.

This perspective underlies the conceptual
and practical advantages of second quantization for 
systems of identical particles and enables a 
unified treatment of matter and electromagnetic 
degrees of freedom.

\subsection{General Structure of the Non-Relativistic QED Hamiltonian}
\label{subsec:nonrel_QED}
By adopting the general framework of QFT, the 
Hamiltonian in Eq.~\eqref{eq:Coulel_Ham} can be 
extended to account for interactions 
between matter and the EMF treated as a quantum dynamical entity\cite{Craig1994,Salam2009}. 
In this formulation, the 
electromagnetic degrees 
of freedom are quantized by applying 
the second quantization procedure to 
the vector potential $\bm{A}(\bm{r},t)$,
from which the observable electric and magnetic fields are obtained as $\bm{E}(\bm{r},t) = -\partial_t \bm{A}(\bm{r},t) - \nabla V(\bm{r},t)$
and $\bm{B}(\bm{r},t) = \nabla \times \bm{A}(\bm{r},t)$.
The redundancy in the description 
introduced by the vector potential is
eliminated by fixing a gauge.
For non-relativistic systems, a 
common and convenient choice is the 
Coulomb gauge where $\nabla \cdot \bm{A}(\bm{r},t) = 0$.
In the Coulomb gauge, it is convenient to expand the 
vector potential in plane-wave modes with wavenumber $\bm{k}$ as
\begin{equation}
\widehat{\bm{A}}(\bm{r},t)=
\sum_{\lambda=1}^2\sum_{\bm{k}}\mathcal{A}(\bm{k})
\left[
\boldsymbol{\epsilon}(\bm{k},\lambda)\, e^{-i\bm{k}\cdot\bm{r}}\,
\widehat{a}_{\bm{k},\lambda}(t)
+
\boldsymbol{\epsilon}^{*}(\bm{k},\lambda)\, e^{i\bm{k}\cdot\bm{r}}\,
\widehat{a}^{\dagger}_{\bm{k},\lambda}(t)
\right],
\end{equation}
where $\mathcal{A}(\bm{k})=[\hbar/(2\varepsilon_0\omega(\bm{k})\mathcal{V})]^{1/2}$
is the quantized field amplitude for a mode of frequency
$\omega(\bm{k})=c|\bm{k}|$ in a finite square box of volume $\mathcal{V}$
(usually introduced to regularize the infrared behavior of the
electromagnetic field), and $\boldsymbol{\epsilon}(\bm{k},\lambda)$ are
the polarization vectors satisfying the transverse condition
$\bm{k}\cdot \boldsymbol{\epsilon}(\bm{k},\lambda)=0$ and the completeness
relation
\[
\sum_{\lambda=1}^{2}
\epsilon^{*}_{i}(\bm{k},\lambda)\,
\epsilon_j(\bm{k},\lambda)
=
\delta_{ij}
-
\frac{k_i k_j}{\|\bm{k}\|^{2}} .
\]
This naturally allows one to define the dyadic 
matrix projector along the transverse directions
$
\delta^{\perp}_{ij}(\bm{r}-\bm{r}')
=
(2\pi)^{-3/2}
\int
\left(
\delta_{ij}
-
\frac{k_i k_j}{\|\bm{k}\|^{2}}
\right)
e^{i\bm{k}\cdot(\bm{r}-\bm{r}')}
\,\mathrm{d}^3\bm{k},$
which is important for identifying the 
dynamical degrees of freedom of the 
electromagnetic field in the Coulomb gauge.

The operators $\widehat{a}^{\dagger}_{\bm{k},\lambda}$ and $\widehat{a}_{\bm{k},\lambda}$ 
are the creation and annihilation operators 
associated with the mode $(\bm{k},\lambda)$, 
acting on the EMF Hilbert's space $\mathcal{H}_{\mathrm{EMF}}$.

Within this representation, the Hamiltonian of the free electromagnetic field reads
\begin{equation}
\widehat{H}_{\mathrm{EMF,free}}[\widehat{a}_{\bm{k},\lambda},\widehat{a}_{\bm{k},\lambda}^{\dagger}]
=\dfrac{\varepsilon_0}{2}
\int
\left[
\|\widehat{\bm{E}}(\bm{r},t)\|^{2}
+
c^2 \|\widehat{\bm{B}}(\bm{r},t)\|^{2}
\right]\,\mathrm{d}^{3}\bm{r}
=
\sum_{\bm{k},\lambda}
\hbar c\,|\bm{k}|\left(
\widehat{n}_{\bm{k},\lambda}+\frac{1}{2}
\right).
\end{equation}
where $c$ is the speed of light. 
The form of the EMF--matter interaction depends on: (i)~the 
representation of matter degrees of freedom (first- vs.\ second-
quantized), (ii)~the relativistic or non-relativistic treatment of 
matter, and (iii)~the coupling formulation (minimal vs.\ 
multipolar).
In the first-quantized framework, the interaction is 
introduced via the \emph{minimal-coupling} prescription: the 
momentum operator of each charged particle is replaced as $
\widehat{\bm{p}}_a \to \widehat{\bm{p}}_a - q\,\widehat{\bm{A}}(\widehat{\bm{r}}_a)$, where $q$ is the particle's charge, in the 
Hamiltonian of Eq.~\eqref{eq:Coulel_Ham}.
In this expression, the vector potential is evaluated at the 
particle  position operator. An equivalent formulation that 
preserves an explicit field-theoretic representation can be 
obtained by expressing the coupling in terms of the electronic current density operator,
$\widehat{\bm{J}}_{\rm el}(\bm{r}) =
-(e/m_e)\sum_{a=1}^{N_e}
\widehat{\bm{p}}_a\,\delta^{3}(\widehat{\bm{r}}_a-\bm{r}),$
together with the electronic charge density operator
$\widehat{\rho}_{\rm el}(\bm{r})$.
The term in minimal-coupling Hamiltonian describing the 
interaction between electrons and the radiative EMF can 
then be written as
\begin{equation}
\label{eq:Ham_elEMF_min}
\begin{split}
&\widehat{H}^{\rm min}_{\rm el\text{--}EMF}[\hat{\bm{r}}_a,\hat{\bm{p}}_a,\widehat{a}_{\bm{k},\lambda},\widehat{a}_{\bm{k},\lambda}^{\dagger}]
=
\sum_{a=1}^{N_e} \frac{e}{m_e}\,
\widehat{\bm{A}}(\widehat{\bm{r}}_a)\!\cdot\! \widehat{\bm{p}}_a
+
\sum_{a=1}^{N_e} \frac{e^2}{2m_e}
\bigl\|\widehat{\bm{A}}(\widehat{\bm{r}}_a)\bigr\|^2
\\
&=
\int \!\left[
\widehat{\bm{J}}_{\rm el}(\bm{r}) \!\cdot\! \widehat{\bm{A}}(\bm{r})
-
\frac{e}{2m_e}\,
\widehat{\rho}_{\rm el}(\bm{r})\,
\bigl\|\widehat{\bm{A}}(\bm{r})\bigr\|^2
\right] \mathrm{d}^3\bm{r}.
\end{split}
\end{equation}
The choice between relativistic and non-relativistic 
treatments of electrons in atomic systems hinges on the ratio of the electron's 
characteristic velocity $v$ in a given spin orbital to the speed of 
light $c$. Relativistic corrections to the energy scale as $(Z v_H/
c)^2$,  and $v_H/c$ denotes the orbital velocity of the electron in 
the ground state of hydrogen — expressible as $v_H/c= a_{\rm Bohr} 
E_{\rm Ha} / (\hbar c)\approx 1/137 $, with $a_{\rm Bohr}$ the Bohr 
radius and $E_{\rm Ha}$ the Hartree energy\cite{dyall2007introduction,reiherRelativisticQuantumChemistry2015}. 
Relativistic quantum mechanics becomes increasingly important for 
accurate core-level energies, and spin–orbit effects while full
relativistic QED treatments are only required for high-precision 
spectroscopic corrections.
However, for atoms with $Z\leq 30$, relativistic corrections to total 
energies are typically at or below the few-percent level, making 
non-relativistic descriptions adequate for many purposes. 
The total nonrelativistic QED (nrQED) 
Hamiltonian 
\begin{equation}
\label{eq:mincoupl_Hint}
\widehat{H}^{\rm min} = \widehat{H}_{\mathrm{EMF,free}}[\widehat{a}_{\bm{k},\lambda},\widehat{a}_{\bm{k},\lambda}^{\dagger}] + \hat H_{\rm el-Coul}[\hat{\bm{r}}_i,\hat{\bm{p}}_i] +\widehat{H}^{\rm min}_{\rm el\text{--}EMF}[\hat{\bm{r}}_i,\hat{\bm{p}}_i,\widehat{a}_{\bm{k},\lambda},\widehat{a}_{\bm{k},\lambda}^{\dagger}]\ ,
\end{equation}
provides a wide applicable and consistent 
description of EMF--matter 
interactions, applicable to both 
extended materials and molecular 
systems. In its minimal-coupling form, 
matter degrees of freedom 
interact locally with the 
electromagnetic vector
potential, making this formulation 
particularly suitable for 
systems with delocalized charges
and translational invariance.

From the structure of Eq.~\eqref{eq:mincoupl_Hint}, 
the coupling between matter and the electromagnetic 
field appears to be set by the electric charge $q$,
as dictated by the minimal-coupling prescription. 
However, the effective strength of this interaction
can be expressed in terms of a dimensionless parameter, $\alpha$. 
In the low-energy regime—where particle--antiparticle creation and 
annihilation processes are negligible—$\alpha$ reduces to the \textit{fine-structure constant},
\begin{equation}
\alpha = \frac{e^{2}}{4\pi\varepsilon_{0}\hbar c} \simeq \frac{1}{137}.
\end{equation}
The smallness of $\alpha$ underlies the success of perturbative 
approaches in low-energy quantum electrodynamics involving a limited 
number of charged particles. By contrast, in systems exhibiting strong 
electronic correlations—arising from high particle densities or specific 
geometric arrangements—effective coupling strengths can be significantly 
enhanced, and non-perturbative matter--EMF effects may emerge, as 
discussed in Sec.~\ref{sec:atmol_inQED}.

For systems composed of spatially localized and 
distinguishable assemblies of bound charges — such as atoms, 
molecules, or molecular aggregates — it is often advantageous 
to adopt an alternative description of the electromagnetic 
field (EMF)–matter interaction, inspired by the multipolar 
framework. 
In this approach, the matter degrees of freedom are 
described not by individual particle coordinates, but by 
macroscopic polarization and magnetization fields. This 
provides a mathematically rigorous formulation of the 
intuitive idea: that confined charge distributions can 
be effectively represented through their multipole 
moments (dipoles, quadrupoles, etc.). This motivates the 
use of the \textit{multipolar coupling} Hamiltonian, 
obtained by applying the Power–Zienau–Woolley (PZW) 
unitary transformation to the combined matter–EMF 
system\cite{Craig1994,Salam2009,woolley2022foundations}. The PZW transformation recasts the EMF–matter
interaction in terms of collective polarization and
magnetization fields associated with spatially 
localized subsystems. For each localized charge set $\xi$,
the interaction is described by the 
polarization field operator
$\widehat{\bm{P}}_{\xi}(\bm{r},t)$, defined such that
$\widehat{\rho}_{\xi}(\bm{r},t) = -\nabla\!\cdot\!\widehat{\bm{P}}_{\xi}(\bm{r},t)$,
together with the magnetization field operator
$\widehat{\bm{M}}_{\xi}(\bm{r},t)$ satisfying $\nabla \times \widehat{\bm{M}}_{\xi}(\bm{r},t)=\partial_t \widehat{\bm{P}}_{\xi}(\bm{r},t)-\hat{\bm{J}}_{\xi}(\bm{r},t)$ and the diamagnetic susceptibility tensor
$\widehat{\bm{O}}_{\xi}(\bm{r},\bm{r}')$
(see Appendix~\ref{sec:Appendix_B}).
The multipolar gauge, introducing a description of the 
quantum electronic degrees of freedom in terms of 
molecular polarization field, provides a rigorous 
and conceptually consistent framework for embedding
and developing approaches to intermolecular non-covalent
dispersion interactions\cite{hunt1983nonlocal, hunt1984nonlocal}.

It is worth noting that the constitutive relations defining 
the polarization and magnetization fields do not uniquely fix 
these fields themselves, but only the associated charge and 
current densities. As a consequence, the polarization and 
magnetization fields are defined only up to transformations 
that leave the physical sources invariant\cite{woolley2022foundations}.
Such a gauge freedom reflects the fact that different 
choices of polarization and magnetization fields can 
represent the same underlying charge distribution. Formally, 
it constitutes the mathematical counterpart of the 
arbitrariness inherent in the choice of reference point (or 
reduction pole) used in the multipole expansion of
an extended electronic charge distribution.

In the multipolar gauge, the nonrelativistic QED Hamiltonian 
can be written as
\begin{equation}
\label{eq:Ham_nrQED_mult}
\begin{split}
&\widehat{H}_{\rm nrQED}^{\rm mult}
=
\widehat{H}_{\rm EMF,free}
+\sum_{\xi}\Biggr\{
\sum_{a_{\xi}\in\xi}
\Bigl[
\widehat{K}_{\rm el}[\widehat{\bm{p}}_{a_{\xi}}]
+\widehat U_{\rm el-el}[\widehat{\bm{r}}_{a_{\xi}}]
+\sum_{A_{\xi}\in \xi}\widehat U_{\rm n-el}[\widehat{\bm{r}}_{a_{\xi}};\widehat{\bm{R}}_{A_{\xi}}]
\Bigr]
+\\[4pt]
&+\dfrac{1}{2\epsilon_0}\int \!\|\widehat{\bm{P}}^{\perp}_{\xi}(\bm{r})\|^{2}\,\mathrm{d}^{3}\bm{r}
-\int \!\left[
\varepsilon_{0}^{-1}\widehat{\bm{P}}^{\perp}_{\xi}(\bm{r})\!\cdot\!\widehat{\bm{D}}^{\perp}(\bm{r})
+\widehat{\bm{B}}(\bm{r})\!\cdot\!\widehat{\bm{M}}_{\xi}(\bm{r})
\right]\mathrm{d}^{3}\bm{r}
+\\
&+\frac{1}{2}\iint
\widehat{\bm{B}}(\bm{r})\!\cdot\!
\widehat{\bm{O}}_{\xi}(\bm{r},\bm{r}')\!\cdot\!
\widehat{\bm{B}}(\bm{r}')\,
\mathrm{d}^{3}\bm{r}\,\mathrm{d}^{3}\bm{r}' \Biggr\}.
\end{split}
\end{equation}
where the displacement vector field $\widehat{\bm{D}}(  \bm{r},t)=\varepsilon_0\widehat{\bm{E}}(\bm{r},t)+\widehat{\bm{P}}(\bm{r},t)$ 
has been introduced, and the 
transverse component $v_i^{\perp}(\bm{r})=\sum_{j}\int \delta_{ij}^{\perp}(\bm{r}-\bm{r}') v_j(\bm{r}')$ of a a vector field $\bm{v}(\bm{r}')\mathrm{d}^3\bm{r}'$ (see Appendix for further details).
Within this formulation, each 
localized matter subsystem 
couples directly only to the 
EMF, while effective matter--matter 
interactions arise from 
field-mediated processes. This structural feature makes the multipolar gauge 
particularly well suited for describing molecular 
spectroscopy, polarization phenomena, and EMF--matter 
interactions in confined or cavity environments.

It is worth emphasizing that, in both the minimal-coupling
 and multipolar formulations, the matter-EMF 
coupling can be rewritten just in terms of the \emph{matter density fields} rather than in terms of individual  particles. 
In a first-quantized description, these 
fields reduce to singular distributions supported 
at the particle coordinates, i.e., Dirac delta 
functions. By contrast, within a second-quantized 
formulation of matter, the electric charge and 
current densities become operator-valued fields 
and admit a mode expansion in terms of the 
underlying matter field operators. The particle-based description is recovered as a particular 
limiting case of this field-theoretic formulation, 
corresponding to states with fixed and pointwise
localized particle content.

In the following sections, we illustrate the 
impact of field-inspired methods in quantum chemistry. We show how second quantization 
provides a natural framework for quantum-information analyses of electronic structure, 
yielding direct insight into covalent bonding 
through field-theoretical observables such as 
the energy–stress tensor. We further review
how effective field-based Hamiltonians enable 
coarse-grained descriptions of electronic 
density and correlations 
(Sec.~\ref{subsec:effFT_description}), 
supporting multiscale modeling and the 
derivation of scaling laws linking 
microscopic electronic correlations to 
macroscopic physico-chemical properties.

\section{Quantum Field Description of Atoms and Molecules}
\label{sec:QFM_section}

This section reviews representative applications 
of field-oriented descriptions of electronic 
quantum matter in chemistry. Subsection 
\ref{subsec:QFM_chembond} revisits the 
Schrödinger field formalism—namely, the 
occupation-number representation of spin orbitals
—in quantum chemistry, with applications to the 
analysis of chemical bonding, including 
approaches based on quantum information
theory. 
Subsection \ref{subsec:effFT_description} 
examines effective field-theoretical models of 
electronic degrees of freedom, with particular 
emphasis on density functional theory and the 
second-quantized formulation of many-body 
dispersion interactions. Finally, subsection 
\ref{subsec:scaling_laws} discusses how field-
oriented formulations of the electronic 
structure problem enable the derivation of 
scaling laws for key physico-chemical 
observables, such as molecular polarizability, 
across a broad range of molecular sizes.

\label{sec:QuantumFieldMatter}
\begin{figure}
    \centering
\includegraphics[width=1\linewidth]{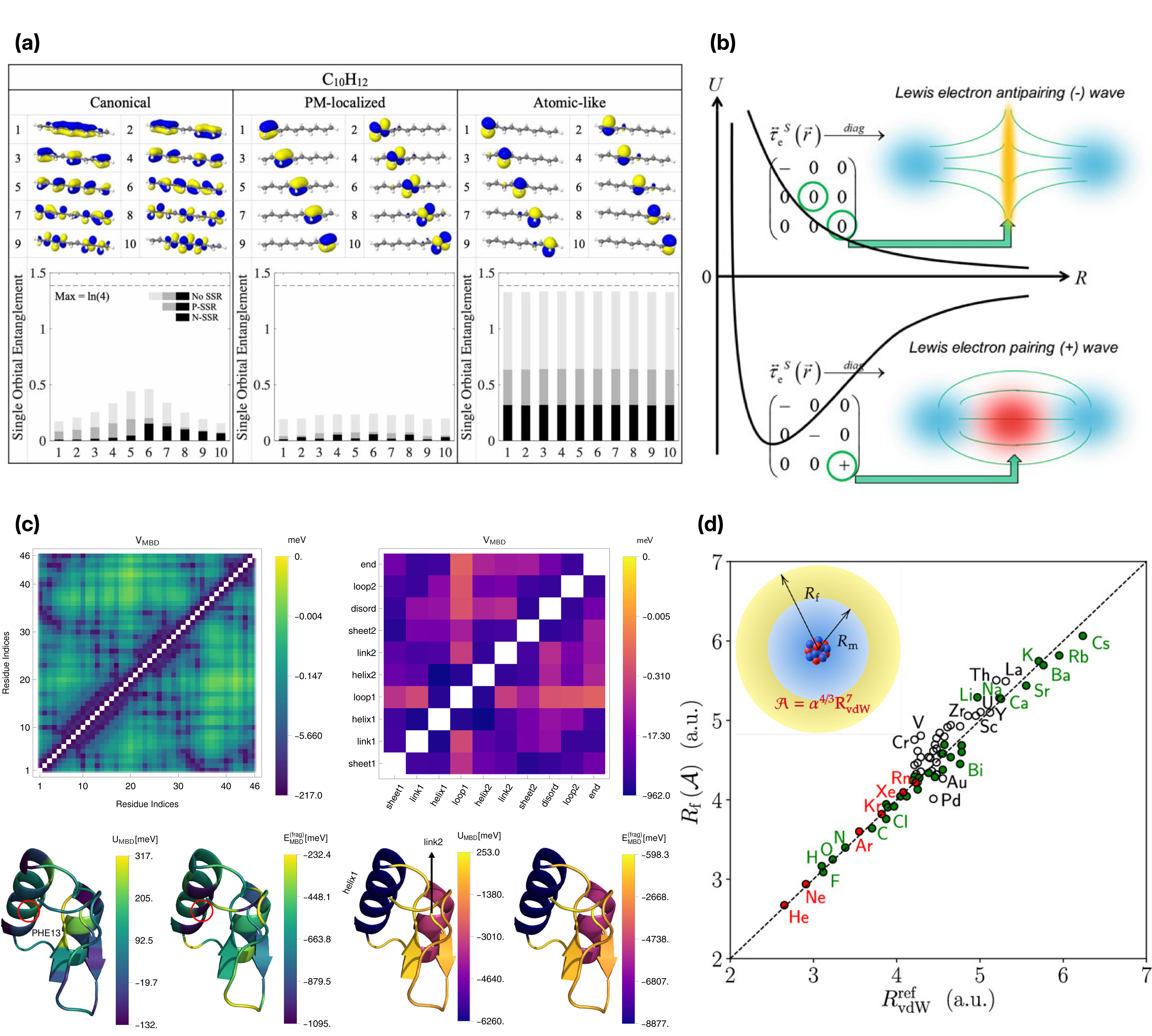}
\caption{\footnotesize \textbf{Applications of QFT-based methods to atomic and molecular systems.}
\textbf{(a)} Single-orbital entanglement (quantum relative entropy) in 
the Complete Active Space calculations 
correlating 10 electrons in 10 $\pi$ 
orbitals for C$_{10}$H$_{12}$. The orbital numbering 
follows the upper panel (canonical 
from self consistent Hartree-Fock/ 
from Pipek-Mezey (PM) localization\cite{pipek1989fast}/
atomic by Jacobi-rotation of PM $\pi$-
orbitals\cite{koridon2021orbital}).
Colors indicate no Superselection Rules (SSR) (all colors), Parity-SSR (black/dark grey), and Number-SSR (black). Adapted from \cite{ding2022quantum}.
\textbf{(b)} Bonding/antibonding characterization in H$_2$-like dimers via eigenvalues and flow lines of the electronic stress tensor $\overleftrightarrow{\tau}_e(\bm r)$: the $1s\sigma$ state exhibits tensile stress ($>0$) at the bond midpoint and a spindle structure in the stress lines, whereas $1s\sigma^\ast$ shows compressive stress ($<0$) and a disrupted stress topology consistent with antibonding character. Adapted from \cite{tachibana2019new}.
\textbf{(c)} Second Quantized Many-Body (SQ-MBD) decomposition of the Many-Body dispersion energy into intra-fragment ($U_{\mathrm{MBD}}$), inter-fragment pair ($V_{\mathrm{MBD}}$), and per-fragment ($E^{\rm frag}_{\rm MBD}$) contributions, shown per residue and secondary-structure element in crambin (meV). Adapted from \cite{gori2023second}.
\textbf{(d)} Reference van der Waals radii $R_{\rm vdW}^{\rm ref}$ \cite{gobre2016efficient} versus electromagnetic field (EMF)-dressed radii for 72 elements, and corresponding polarizabilities (in atomic units); noble gases (red), transition metals (open symbols), and other elements (green). The plot illustrates the scaling relation between static polarizability $\mathcal{A}$ and the EMF-dressed van der Waals radius $R_{\rm f}$, with a prefactor involving the fine-structure constant $\alpha$. Adapted from \cite{tkatchenko2021fine}.}
\label{fig:Application_QMF}
\end{figure}

\subsection{Field-Inspired Approaches to Matter Hamiltonian and Chemical Bonding}
\label{subsec:QFM_chembond}
Different possible QFT representations of
the electronic degrees of freedom have been applied
in the context of quantum chemistry, depending on 
the desired level of accuracy (e.g., whether 
relativistic effects must be included):
4-component Dirac spinor field\cite{Koch2025-Relativistic-QED},
Schr\"odinger wavefunction field, or charge-density field.
The Schr\"odinger field representation of the electronic
degrees of freedom is the closest quantum field 
theoretical approach to the usual wavefunction
methods.\cite{joergensen2012second,surjan2012second}
The second quantization scheme presented in the
previous section can be applied to 
the Schr\"odinger field, 
where the electronic degrees of  
freedom are expressed in terms of orthonormal spin orbitals,
$\boldsymbol{\phi}_{s,\eta}(\bm{r}) = \phi_{\eta}(\bm{r})\,\mathbf{v}_s$,
where $\{\phi_{\eta}(\bm{r})\}_{\eta=1,\ldots,+\infty}$
denotes a chosen complete basis set of 
square-integrable functions on a domain 
in real space while $\mathbf{v}_{\uparrow} = (1,0)^{T}$
and $\mathbf{v}_{\downarrow} = (0,1)^{T}$ are the
spin vectors.
The corresponding (adjoint) field operator is
\begin{equation}
\widehat{\boldsymbol{\psi}}^{(\dagger)}(\bm{r}) =
\sum_{s=\uparrow,\downarrow} \widehat{\boldsymbol{\psi}}_{s}^{(\dagger)}(\bm{r})= \sum_{\eta \in \mathcal{O}^{\mathrm{mol}}}
\boldsymbol{\phi}_{s,\eta}^{(*)}(\bm{r})\,\widehat{c}^{(\dagger)}_{\eta,s},
\end{equation}
where $\widehat{c}^{\dagger}_{\eta,s}$ and $\widehat{c}_{\eta,s}$ are the 
fermionic creation and annihilation 
operators for the spin orbital labeled by $(\eta,s)$.
Acting on Fock space, these operators generate
many-electron basis states of the form
$\lvert n_{\eta_1,s_1}, \ldots, n_{\eta_k,s_k}, \ldots \rangle$, labelled by 
the occupation number of each spin orbital.
In such a QFT picture, the electric 
charge and current density 
distributions read
\begin{align}
&\widehat{\rho}_{\mathrm{el}}(\bm{r}) =-e \sum_{s=\uparrow,\downarrow} \widehat{\psi}^{\dagger}_{s}(\bm{r}) \widehat{\psi}_{s}(\bm{r})=-e\sum_{s=\uparrow,\downarrow}\sum_{\eta,\eta'\in\mathcal{O}^{\rm mol}}\phi^{*}_{\eta'}(\bm{r}) \phi_{\eta}(\bm{r}) \hat{c}_{\eta',s}^{\dagger}\hat{c}_{\eta,s}\\
&\widehat{\bm{J}}_{\mathrm{el}}(\bm{r}) = \dfrac{ie\hbar}{2m_e}\sum_{s=\uparrow,\downarrow} \left[\widehat{\psi}^{\dagger}_{s}(\bm{r}) \nabla_{\bm{r}}\widehat{\psi}_{s}(\bm{r})-\left(\nabla_{\bm{r}}\widehat{\psi}^{\dagger}_{s}(\bm{r})\right) \widehat{\psi}_{s}(\bm{r})\right]=\\
\nonumber &=\dfrac{ie\hbar}{2m_e}\sum_{s=\uparrow,\downarrow}\sum_{\eta,\eta'\in\mathcal{O}^{\rm mol}}\left[ \phi^{*}_{\eta'}(\bm{r})\nabla\phi_{\eta}(\bm{r}) - \left(\nabla\phi^{*}_{\eta'}(\bm{r})\right)\phi_{\eta}(\bm{r})\right] \hat{c}_{\eta',s}^{\dagger}\hat{c}_{\eta,s}\,\,.
\end{align}

Such a second-quantized description of the
nonrelativistic electronic wave function forms the
foundation of modern electronic structure methods, 
including
Hartree--Fock theory and the coupled-cluster (CC)
approach\cite{bartlett2007coupled,faulstich2024recent}, for which second quantization constitutes a fundamental basis.
Recently, renewed interest in field-based
formulations of CC theory has highlighted the
connections to algebraic
geometry\cite{faulstich2024coupled,faulstich2024algebraic}, offering new perspectives on the 
structure and solvability of the 
coupled-cluster equations.
Moreover, the second-quantization formulation of the electronic Hamiltonian in quantum chemistry
enables the determination of ground-state properties using quantum computational 
methods\cite{kassal2011simulating,whitfield2011simulation,ryabinkin2018qubit,bauer2020quantum,patel2025quantum}. 
In particular, the Jordan–Wigner
transformation provides a mapping from fermionic modes to qubit operators, thereby allowing
the quantum chemistry Hamiltonian to be expressed as a sum of qubit operations.

Recent research highlights the second-quantized framework 
as a natural, powerful setting for applying quantum 
information tools to quantum chemistry. 
Central to this is the concept of \textit{orbital 
entanglement}~\cite{boguslawski2015orbital,friis2016reasonable,ding2020concept,ding2022quantum,aliverti2024can}: in 
identical-particle systems, subsystems are defined not by 
partitioning particles, but by partitioning field modes --- e.g., electronic spin-orbitals.
Operators for a single fermion (e.g., an electron) 
do not form a valid subsystem in many-body fermionic systems~\cite{aliverti2024can} from which it follows
that first-quantized RDMs --- obtained by tracing out 
electron positions --- are conceptually problematic. In 
contrast, the algebra of creation/annihilation operators 
for a single orbital defines a proper subsystem, enabling 
consistent 1- and 2-orbital RDMs. Some care is required 
in dealing with fermionic RDMs, as creation and 
annihilation operators for different orbitals do not 
commute in general — a feature consistent with the 
microcausality principle in non-relativistic settings~\cite{johansson2016comment}. 
This implies that superselection rules 
must be applied to the RDM, forbidding 
superpositions of fermionic states with different 
particle-number parity — the so-called parity 
superselection rule (P-SSR) — or preserving 
the total number of fermions — the number superselection rule 
(N-SSR)\cite{friis2016reasonable,ding2020concept}.
Within this framework, quantum information methods
yield new insights --- e.g., via \textit{orbital} 
entanglement estimators applied across spin-orbital 
bases to identify those best describing covalent 
bonds, bridging molecular orbital and valence bond 
theories. Atomic orbital bases maximizing orbital 
entanglement estimators are most suited (see 
panel (a) of Figure~\ref{fig:Application_QMF}) for 
characterizing covalent bonds.
Based on this, a new method assesses active orbital 
space quality in 
Complete Active Space Configuration Interaction~\cite{ding2023quantum,liao2024quantum}, 
yielding results analogous to first-quantized CC 
via Domain-Based Local Pair Natural Orbital~\cite{riplinger2013efficient,schneider2016decomposition}.

The field-theoretical approaches for 
describing electronic quantum states 
allow one to adapt methods derived from
continuous 
mechanics and quantum fluid
mechanics to gain a deeper understanding of 
chemical bonds. A remarkable example in this sense
is the ``rigged'' QED formalism for 
describing covalent and hydrogen bonding
(see \cite{tachibana2004spindle}\cite{tachibana2017new}
and references therein for a comprehensive 
perspective).
In this approach —particularly within the
\textit{primary} rigged QED framework
tailored for quantum chemistry — both electrons 
and nuclei are treated as quantum Schr\"odinger 
fields coupled to the quantized electromagnetic 
field. Evaluating these fields in the 
ground state yields classical field 
representations, 
from which the real-space energy projection 
— the system’s rank-2 stress–energy tensor 
— is derived.
Analyzing the flow lines of
its principal eigenvector reveals a 
characteristic spindle-shaped pattern signature 
of covalent bonding (see panel (b) of Figure \ref{fig:Application_QMF}).
Recent extensions have captured subtle features 
of hydrogen bonding \cite{senami2020identification}, 
underscoring the power of quantum field-based 
descriptions of matter to uncover the field-theoretic 
roots of molecular interactions.
Field-inspired approaches are effective 
in describing the  real-space embedding of 
electronic structure properties beyond the local scale of 
covalent bonds.

Because position in real space functions as a label 
rather than a quantum variable, field-theoretic 
formulations inherently span all length scales. When 
applied to the electronic-structure problem in 
Eq.~\eqref{eq:Coulel_Ham}, these approaches provide new 
insight into electronic correlation, particularly in the 
treatment of long-range, non-covalent interactions.

\subsection{Non-Covalent Interactions from 
Effective Field Models for Electronic Density and Polarization} 
\label{subsec:effFT_description}

A field-theoretic formulation of electronic degrees of freedom grants direct access to real-space densities of physical observables, enabling spatially resolved characterization of intra- and intermolecular interactions. Central among these is the electronic charge density
which underpins most molecular properties. The electron density is both 
computationally tractable and experimentally measurable --- 
notably via X-ray diffraction and quantum tomography --- 
and its geometric and topological features have been 
systematically analyzed to decode chemical bonding 
mechanisms~\cite{martinpendasTopologicalApproachesChemical2023,koch2024analysis}.
The centrality of the expectation value of the 
charge density field $\rho_{\mathrm{el}}(\mathbf{r}) = \langle \widehat{\rho}_{\mathrm{el}}(\mathbf{r}) \rangle_{\rm GS}$ in the ground state is formalized by the Hohenberg--Kohn theorem: a one-to-one mapping exists between the nuclear potential $\widehat{U}_{\rm n\text{-}el}(\mathbf{r})$ and $\rho_{\mathrm{el}}(\mathbf{r})$, derived from the Hamiltonian in Eq.~\eqref{eq:Coulel_Ham}. This implies an energy functional $
E[\rho_{\mathrm{el}}] = F[\rho_{\mathrm{el}}] + U_{\mathrm{el\text{-}n}}[\rho_{\mathrm{el}}],$  
where $F[\rho_{\mathrm{el}}] = T_s[\rho_{\mathrm{el}}] + E_{\rm H}[\rho_{\mathrm{el}}] + E_{\rm xc}[\rho_{\mathrm{el}}]$ is universal --- encoding kinetic, Hartree, and exchange--correlation contributions --- yet remains unknown in exact form.
In Kohn--Sham DFT, the exchange-correlation functional $E_{\rm xc}[\rho_{\mathrm{el}}]$, in principle, accounts for all beyond-Hartree correlations. 
Its form, especially for what concerns long-range component --- critical for nonlocal 
effects --- remains the principal modeling challenge, 
motivating advanced approximations. QFT methods provide a first-principles route to 
$F[\rho_{\mathrm{el}}]$: DFT arises naturally as an 
effective low-energy field theory for the density 
operator~\cite{polonyi2002effective,fernando2008generalized}--in the sense of 
an effective action for 
the density -- enabling 
perturbative 
constructions of the 
exchange-correlation 
functional $E_{\rm xc}$.
Non-perturbative approaches, inspired 
by functional renormalization, have recently yielded 
3D universal functionals~\cite{yokota2021ab}. Despite decades of progress in constructing 
increasingly accurate energy density functionals, 
no universally applicable functional yet exists that 
reliably captures the contribution of electronic 
density correlations to interaction energies --- 
particularly in non-covalent, many-body regimes. 
A promising strategy consists in defining model 
integrable Hamiltonians encoding electronic 
density correlations via physically motivated 
representations.
A prominent example is the Many Body 
Dispersion (MBD) 
model~\cite{tkatchenko2012accurate,ambrosetti2014long,hermann2017first,stohr2019theory}. 
In this  framework, the electronic 
response --- derived from a 
chosen density functional approximation 
(DFA) --- is mapped onto a system of quantum
Drude oscillators (QDOs), which interact via
a dipole--dipole electrostatic potential. 
The MBD Hamiltonian thus provides a low-energy 
effective description of long-range
electronic correlations, typically manifesting at
length scales $\gg 5\,\text{\AA}$, through the 
collective normal modes of the coupled QDO system.
This formalism yields computationally efficient, 
physically transparent expressions for the MBD energy --- a contribution 
systematically absent in semi-local DFAs --- 
enabling systematic studies of dispersion effects 
in large molecular complexes and extended 
materials.  Moreover, MBD has been validated as a 
quantitative proxy for charge density distortion 
relative to high-accuracy references (e.g., 
CCSD(T)), when initialized with DFA-derived polarizabilities\cite{khabibrakhmanov2025noncovalent}.
A second-quantized formulation of MBD, inspired by quantum 
field theory, was recently introduced~\cite{gori2023second}. 
It maps collective MBD modes --- and the ground state of the 
interacting QDO system --- onto atomic QDO displacements and 
non-interacting eigenstates via a Bogoliubov transformation. 
This allows one to derive a fragment-resolved 
decomposition of the MBD energy (see panel (c) of
Figure~\ref{fig:Application_QMF}), 
and provides a rigorous basis for applying quantum 
information tools to 
quantify correlation pathways in the MBD-
induced charge density~\cite{gori2023second}.

These examples indicate that a field-based formulation 
offers a natural and systematic framework for the 
construction of effective Hamiltonians describing matter 
degrees of freedom across different length and energy 
scales. In this context, a compelling extension of the many-body dispersion (MBD) framework is to introduce an effective 
Hamiltonian for quantum electronic charge-density 
fluctuations defined with respect to a reference 
semiclassical electronic density obtained from density-
functional theory (DFT). Conceptually, this approach 
parallels semiclassical approximations in the path-integral 
formulation of quantum field theory, where the dynamics of 
fluctuations are treated perturbatively around a saddle-
point configuration.
This formulation would transcend the current quantum Drude 
oscillator (QDO) paradigm by enabling an optimized and self-
consistent determination of the spatial distribution, 
parametrization, and coupling constants of the effective 
oscillators, rather than prescribing them through semi-
empirical models.

Field-inspired approaches provide a natural 
bridge between atomistic and molecular length 
scales. The systematic connection across these 
scales is most effectively captured through the 
derivation of scaling laws — quantitative 
relationships that encode how 
physical properties evolve with system size, number of particles, composition, or geometry.

\subsection{Scaling Laws for Covalent and Non-Covalent Interactions from Field Theory Perspective}
\label{subsec:scaling_laws}


A central limitation of first-quantized quantum 
mechanics in the study of scaling laws
is its reliance on a fixed-particle-number description. 
When the number of electrons
changes across systems—an unavoidable situation when 
exploring chemical space—scaling
relations for quantum-mechanical observables such as 
energies, multipole moments, or 
polarizabilities cannot be formulated in a unified 
manner.
By contrast, QFT-inspired 
approaches operate
within
an occupation-number 
representation, naturally 
embedding Hilbert spaces corresponding to different particle numbers and thereby providing a consistent
framework for analyzing trends across families of molecular systems.

This perspective underlies field-theoretic and field-
inspired approaches to chemical bonding. A notable
example is Alchemical Perturbation Density Functional 
Theory (APDFT), which exploits functional derivatives of 
the ground-state DFT energy and electron
density with respect to the nuclear density to define an \textit{alchemical potential}
\cite{von2005variational,von2020alchemical}.
This observable quantifies the response of
electronic structure to changes in nuclear charge and 
provides chemically meaningful insight into bonding 
trends, such as variations in acidity or reactivity. By 
enabling the exploration of entire families of 
isoelectronic compounds without explicit enumeration, 
APDFT illustrates how field-theoretic concepts allow 
systematic navigation of chemical space.

Field-theoretic methods 
provide a natural route to 
derive scaling laws for 
extensive response 
properties, due to 
flexibility of 
these methods in 
describing systems
with different number of particles.
In particular, recent work has shown
that the static polarizability $\mathcal{A}$ of systems 
governed by central potentials scales as $\mathcal{A}\propto L^4/a_0$, where $L$ is the characteristic length 
scale
of the ground-state charge density and $a_0$ the Bohr 
radius, rather than following a
naive volume scaling \cite{szabo2022four,goger2024four}. 
This result underpins the
theoretical derivation of the empirical relation
$R_{\rm vdW} \sim \mathcal{A}^{1/7}$ for van der Waals radii \cite{tkatchenko2021fine} (see panel (d) of Figure~\ref{fig:Application_QMF}).
Importantly, the proportionality constant in this scaling 
law encodes the fine-structure
constant $\alpha$, revealing that van der Waals radii 
originate from a quantum electrodynamical dressing of electronic degrees of freedom by low-energy virtual photons.

These examples show that scaling laws for both 
covalent and non-covalent interactions arise 
naturally when the electronic structure is formulated 
within a field-theoretic framework. A central 
challenge moving forward is the development of fully 
first-principles, QFT-based approaches capable of 
deriving scaling relations for a broad range of 
observables across molecular systems spanning vastly 
different length scales. A particularly important 
direction is the extension of scaling laws linking 
the real-space geometry of the electronic charge 
density to extensive response properties, such as 
the static polarizability. These properties govern 
long-range interactions and collective phenomena in 
large molecular assemblies, including biomolecular 
complexes comprising millions of atoms in realistic 
aqueous environments. Understanding how electronic 
and nuclear charge distributions in real space 
determine electronic-structure properties is 
therefore essential to develop accurate coarse-grained
 interaction models and to build 
chemical fragment databases to train machine-learning force fields with quantum-chemical accuracy\cite{unke2024biomolecular,kabylda2025molecular}.

The results presented in this section 
demonstrate the power of the QFT formalism and techniques in 
providing a unified understanding of quantum 
chemistry, as described by 
Eq.~\eqref{eq:Coulel_Ham}, across multiple 
length scales. However, as systems are scaled 
to larger and more complex structures, 
coupling to the quantized radiative 
electromagnetic field can no longer be 
neglected. In the next section, 
we discuss quantum-chemical 
applications in which treating the 
electromagnetic field as a dynamical quantum 
degree of freedom is essential for 
reproducing experimental observations. For 
this purpose it is necessary to consider the 
framework of non-relativistic QED where 
matter and radiative dynamical EMF fields are 
mutually coupled.

\section{Atoms and Molecules Coupled with Quantum Fields}
\label{sec:atmol_inQED}
Atoms and molecules are constantly subject to quantum vacuum fluctuations arising from all underlying fields, mainly the electromagnetic field, but also the Dirac electron-positron field. The fluctuating electromagnetic field (EMF) interacts with matter and can lead to feedback mechanisms, whereby the field alters the properties of matter and matter responds by exciting the field. The strength of these interactions is determined by how the field–matter coupling compares with the relevant loss and decoherence rates in both subsystems, \textit{i.e.}, relaxation and dephasing of the material excitations and photon loss of electromagnetic modes. When the rate of coherent field–matter energy exchange is slower than these decay processes, the system is in the weak-coupling regime. Such interactions can be studied perturbatively by treating them as perturbations to the isolated field and matter systems. In contrast, when the energy exchange between field and matter becomes faster than the losses of the combined system, the matter-field system must be described by hybridized states~\cite{Haroche2006-exploring-the-quantum}, which can decouple after some time (\textit{e.g.}, in a cavity). Hence, the dynamics of each entity in the combined system become strongly interdependent, leading to hybrid states known as polaritons. The rapid, repetitive exchange of energy that occurs before energy leaks out of the system is called Rabi oscillation. Instead of the molecule's original energy levels, the hybrid state splits into two distinct levels (Upper and Lower Polaritons), separated by a gap called the Rabi frequency~\cite{Ebbesen2016-Perspective-Hybrid-Light-Matter, Huo2023-Review-PolaritonChemistry}. By combining quantum mechanical descriptions of the electromagnetic field, for example, as found in cavity quantum electrodynamics, with the complete molecular characterization based on potential energy surfaces used in chemistry, these hybrid states are studied. In such studies, the goal is to generalize well-developed concepts from quantum chemistry to field-matter hybrids.

The realization that fluctuating electromagnetic fields mediate molecular interactions has revealed new opportunities to tailor material behavior by controlling those fluctuations through spatial confinement or external fields. The electronic structure and properties of the molecules ``dressed'' by the fluctuating electromagnetic field in general deviate from those of the isolated molecules. 
To understand the physical concept of a dressed state, one can consider the paradigmatic shift in the energy levels of a hydrogen atom known as the Lamb shift. Although conceptually distinct, both the Lamb shift and polaritonic level restructuring originate from the coupling of matter to the quantized electromagnetic field and can be viewed as manifestations of electromagnetic dressing in different coupling regimes.
Understanding such states requires a theoretical framework in which both the field and matter degrees of freedom are treated quantum-mechanically and evolve under the same fundamental principles.  
In this section, we discuss non-covalent and covalent interactions in QED and explain why a quantum electrodynamical extension of conventional quantum chemistry is essential for studying the mediation and modification of these molecular interactions via vacuum fluctuations.

\subsection{Weak Matter-Field Coupling}
Weak matter-field coupling is the fundamental mechanism behind numerous physical phenomena, including the Lamb shift and spontaneous emission. In molecular QED, intermolecular interactions are mediated through the vacuum electromagnetic field. Resonant energy transfer, vdW dispersion interactions, and scattering phenomena are examples of field-mediated interactions between atomic and molecular entities. In these cases, matter excitations are transferred to the electromagnetic field with probabilities that depend on the field’s density of states. Consequently, they are influenced by the environment through its impact on the density of states of the field. Therefore, an advantage of considering these interactions from the QED perspective is that any change in the environment in which the systems interact, or the application of external fields to atoms and molecules, can be accounted for by examining how these changes affect fluctuations of the vacuum electromagnetic field.
For example, imposing macroscopic boundaries on a system induces reflections of the vacuum field from these boundaries, thus altering the distribution of the field’s modes and, consequently, modifying the molecular interactions. Such scenarios are the focus of macroscopic QED~\cite{Buhmann-Disp-I-2013}.

On the other hand, variations in matter systems, particularly changes in the electric and magnetic properties of atoms or molecules, can also alter how these entities interact with the vacuum field, thereby influencing molecular interactions. For instance, since chiral molecules exhibit distinct interactions with electromagnetic fields compared to non-chiral ones, molecular interactions within chiral systems also differ. Another example is when interacting atoms are initially excited. In such cases, in addition to the vdW dispersion interaction, resonance contribution is also present due to the exchange of real photons between the molecules~\cite{myPRL_2015, Power-PRA1995-dispersion-excited-molecules}.

To evaluate the interaction energy between atoms, one must solve the equation of motion for the entire matter-field system, which is challenging even in the simplest scenario of two interacting atoms.
Consequently, approximate methods and simplified physical models are necessary to address this problem.
Among the approximate approaches in the weak coupling regime, perturbation theory stands out as one of the most widely employed, within which atom-field and atom-atom couplings are considered perturbatively. However, this approach is only practical for a small number of atoms, thereby avoiding higher-order perturbations. In the case of two atoms, the states of the unperturbed system with $\opH_0=\opH_{\rm atom}+\opH_{\rm field}$ (see section \ref{subsec:nonrel_QED} and Appendix~\ref{sec:Appendix_B} for the definition of $\opH_{\rm atom}$ and $\opH_{\rm field}$ in the minimal-coupling formalism of QED) are expressed as the product states of the atoms and the field, {\it i.e.}
\begin{equation}
|\Psi^{(0)}\rangle=|\psi^{(0)}(\bmr_1)\rangle |\psi^{(0)}(\bmr_2)\rangle |n_{k_1\lambda_1},n_{k_2\lambda_2, \cdots}\rangle \ ,
\end{equation}
where $n_{k_i\lambda_i}$ denote the occupation numbers of the field's modes with frequency $\omega_i=ck_i$ and polarization $\lambda_i$, and $|\psi^{(0)}(\bmr_i)\rangle$ are the eigenkets of isolated atoms. Using these states, perturbation calculations can be performed to determine the energy of the total system under the perturbation Hamiltonian~\eqref{eq:Ham_elEMF_min}.
As is well-known in quantum field theory and QED, the energy of the matter-field system is infinite due to the fact that the energy of the fluctuating field in its vacuum state is the sum of zero-point energy of infinity of modes, $E_{\rm f}^{(0)}=\sum_{i=1}^{\infty}\hbar ck_i$.
Nevertheless, it is evident that the interaction energy between the two atoms, which is dependent on the interatomic distance $R$, is finite and can be obtained by renormalizing the energy of the interacting matter-field system with respect to the energy of the non-interacting system $(R\to\infty)$.

An elegant approach to perturbative evaluation of interaction energies in QED and quantum field theory is the use of Feynman diagrams. In the minimal-coupling formalism of molecular QED, typical Feynman diagrams representing the interaction of two atoms resemble those depicted in figure~\ref{fig:dispersion-int}~\cite{Salam2016-nonrQED}. Vertices in these diagrams denote atom-atom or atom-field interactions accompanied by virtual transitions of the subsystems, thereby indicating the perturbation order for each scenario depicted in the diagram. Consequently, diagrams $(i)$ and $(ii)$ represent second-order perturbation, $(iii)$ and $(iv)$ represent third-order perturbation, and diagram $(v)$ represents fourth-order perturbation. Diagram $(i)$ illustrates an interaction between the atoms from second-order perturbation with the instantaneous dipolar Coulomb coupling $V_{\rm dd}$. This diagram corresponds to the same physical process responsible for London dispersion interaction in semi-classical theory.
In contrast, in diagram $(ii)$, the interaction is solely attributed to the term $\opA^2$ and the exchange of two virtual photons that are emitted or absorbed simultaneously at the atomic centers.
The third-order diagrams $(iii)$ and $(iv)$ depict interactions arising from mixed processes. In diagram $(iii)$, an instantaneous dipolar Coulomb coupling ($V_{\rm dd}$) is followed by the exchange of a virtual photon between the atoms due to two consecutive atom-field couplings $\opp\cdot\opA$ at the two atomic centers.
The physical process underlying diagram $(iv)$ involves the exchange of two virtual photons between the two atoms. In the first atom-field coupling ($\opp\cdot\opA$) at the atomic center $B$, a virtual photon is emitted. Subsequently, atom $A$ interacts with the field via $\opA^2$, which involves simultaneous emission and absorption of two virtual photons [Note that $\opA^2$ includes combinations of creation and annihilation operators of photons ($\opa^\dagger_n$ and $\opa_n$, respectiveluy) such as $\opa^\dagger_n \opa_n$ and $\opa_n\opa^\dagger_n$, enabling two-photon processes through $\opA^2$ atom-field couplings. For more details about these bosonic operators and their commutation relations, refer to Appendix~\ref {sec:Appendix_A}]. Finally, the remaining photon is absorbed in another $\opp\cdot\opA$ interaction at atom $B$.
The fourth-order diagram $(v)$ is the result of four consecutive $\opp\cdot\opA$ interactions (two at each atom), which lead to the exchange of two virtual photons between the atoms.

Since the dipole polarizability of an atom is proportional to $e^2$, or equivalently to the dimensionless fine-structure constant $\alpha$, the power of $e^2$ or $\alpha$ can be regarded as a measure of the magnitude of atom-atom and atom-field interactions. Considering the order of perturbations and the physical processes associated with each diagram in figure~\ref{fig:dispersion-int}, it is evident that all five diagrams exhibit the same power of $e^4$ (or $\alpha^2$), indicating that they possess comparable physical significance, despite originating from different perturbation orders. Naturally, there are additional diagrams, but all of them can be classified into one of these five categories. By summing over all possible diagrams, the total interaction energy between the two atoms can be determined.
Including all possible diagrams is essential to ensure that the interaction energy is computed from a physical picture consistent with QED principles. In particular, in diagrams $(i)$ and $(iii)$, the two atoms interact via an instantaneous dipolar Coulomb coupling, which, at first, might seem unphysical within the QED framework. However, considering every relevant contribution of the same order with respect to the fine structure constant ($\alpha$), all instantaneous contributions to the interaction energy cancel out~\cite{Casimir_Polder1948}, and the remaining expression is fully retarded, thereby complying with the relativistic nature of the fluctuating electromagnetic field. This cancellation of instantaneous interactions indicates that, in molecular QED, all interactions are effectively mediated through the field and the exchange of transverse virtual photons.

The mediation of intermolecular interactions by the fluctuating electromagnetic field explains the retardation effects on these interactions. For two atoms to interact, they must interact with the field and borrow polarization and energy from it in the form of virtual transitions. Then, the borrowed polarization and energies enable the atoms to interact via the exchange of virtual photons. However, the borrowing process is subject to the uncertainty principle. The borrowed energy must be returned within a time frame that does not violate the energy-time uncertainty. This means that, at large interatomic distances, where virtual photons have to travel for a long time, those virtual photons belonging to the electromagnetic field’s modes with higher frequencies contribute less compared to those with lower frequencies. Conversely, high-frequency modes contribute more significantly to the interaction when the interatomic distance is small~\cite{Craig1994}.
In the non-retarded limit, where the interatomic distance $R$ is significantly smaller than the characteristic atomic wavelengths, i.e., $R \ll \lambda$, the general expression for the interaction energy can be approximated to reproduce the London dispersion interaction ($R^{-6}$). In the opposite limit, where $R \gg \lambda$, the interaction energy is given by the Casimir-Polder expression ($R^{-7}$)~\cite{Casimir_Polder1948}.

An alternative formalism of molecular QED can be obtained by applying a quantum canonical transformation, known as the Power-Zienau-Woolley transformation, to the minimal-coupling Hamiltonian~\eqref{eq:Ham_elEMF_min}. As it is shown in Appendix~\ref{sec:Appendix_B}, in the resulting Hamiltonian, matter-field interactions arise from the coupling of matter's electric and magnetic multipole moments to the electric and magnetic fields; hence, the formalism is referred to as multipolar coupling. In dipole approximation, the interaction Hamiltonian is given by $\widehat{H}_{\rm int}^{\rm mult} = - \sum_\xi \epsilon_0^{-1} \widehat{\bm{\mu}}_\xi \cdot \widehat{\bm{D}}^\perp(\bmR_\xi)$, where $\widehat{\bm{\mu}}_\xi$ is the electric dipole moment of atom $\xi$ and $\widehat{\bm{D}}^\perp$ is the transverse component of the microscopic displacement field (there is also another term in the multipolar Hamiltonian that must be taken into account when computing self-energies, {\it e.g.} Lamb shift, but does not play a role in the interaction between atoms or molecules~\cite{Craig1994, Salam2009}). In this formalism, the interaction Hamiltonian consists solely of couplings between matter and the field, with no direct couplings among the matter components; therefore, the retarded nature of the interactions is evident from the Hamiltonian. Additionally, the multipolar form of matter-field interactions facilitates the interpretation of the physical behavior of the resulting interaction energies, as they have counterparts in classical electromagnetism.
It is seen that in multipolar formalism, the dispersion interaction between two neutral nonpolar atoms is obtained from fourth-order perturbation theory with the interaction Hamiltonian $\widehat{H}_{\rm int}^{\rm mult}$~\cite{Craig1994, Salam2009}.

As noted earlier, one advantage of the QED perspective on intermolecular interactions is that it enables us to understand how physical conditions influence these interactions by considering their effects on the quantum-mechanical fluctuations of the electromagnetic field. Introducing media or boundaries, such as macroscopic bodies, disrupts and alters these fluctuations. Scattering of the electromagnetic field at these boundaries alters the distribution of modes and their fluctuations, leading to new interactions between atoms. In macroscopic QED, these medium-induced effects are analyzed using classical Green’s functions that incorporate all the geometric and electromagnetic properties of the environment in which the atoms interact~\cite{Buhmann-Disp-I-2013}.

An alternative approach to modifying the fluctuating electromagnetic field is to apply an external dynamic field, such as a monochromatic laser beam. This external field can be regarded as an excitation of the vacuum field, thereby producing real photons in addition to the virtual photons that atoms use for interaction. In an intense field, dispersion interactions undergo significant changes and scale with the interatomic distance $R$, unlike the well-known Casimir-Polder and London dispersion energies. These changes depend on the polarization of the applied field and the orientation of the interacting pair relative to the direction of the field’s propagation. When averaged over all directions of $\bmR$, an attractive dispersion interaction $\propto R^{-1}$ is observed in the near zone. Conversely, in the far zone, the dispersion interaction is proportional to $R^{-2}k_{\rm f}\sin(2k_{\rm f}R)$, where $ck_{\rm f}$ represents the frequency of the applied field. Consequently, the interaction in the far zone can be either attractive or repulsive depending on the ratio $k_{\rm f}/R$~\cite{Thirunamachandran1980-external-field}.

Instead, if the external field is static, the dynamics of the electromagnetic field remain unchanged, but the atoms acquire additional static dipole moments that allow new channels of interaction with the vacuum field and with one another. In the multipolar-coupling scheme, this corresponds to extending the interaction Hamiltonian to 
$\widehat{H}_{\rm int}^{\rm mult} = - \sum_\xi \epsilon_0^{-1} [\widehat{\bm{\mu}}_\xi^{(f)}+\widehat{\bm{\mu}}_\xi^{(s)}] \cdot \widehat{\bm{D}}^\perp(\bmR_\xi)$, 
where $\widehat{\bm{\mu}}_\xi^{(s)}$ is the external-field–induced dipole and $\widehat{\bm{\mu}}_\xi^{(f)}$ the intrinsic fluctuating dipole. Perturbation theory then yields interaction-energy contributions 
$\Delta E= \Delta E^{(2)}_{ss}+\Delta E^{(4)}_{fs} + \Delta E^{(4)}_{ff}+\cdots\ $, 
where each term corresponds to a distinct atom–field coupling channel. The 
$ss$ channel describes the interaction of two induced dipoles mediated by the vacuum field, giving a field-induced electrostatic term $\propto R^{-3}$ that may be attractive or repulsive. The mixed $fs$ channel corresponds to the coupling of a fluctuating dipole on one atom with an induced dipole on the other, producing an always-attractive polarization interaction $\propto R^{-6}$. Finally, the $ff$ channel represents the coupling of two fluctuating dipoles and produces the familiar dispersion interaction $\propto R^{-6}$ (nonretarded) or $\propto R^{-7}$ (retarded). The many-body character of these forces and the interplay between them imply that external static fields can be used to tailor noncovalent interactions (for example, in a benzene dimer as shown in Fig.~\ref{fig:external-field})~\cite{Reza_PRR2022, Reza_JPCL2022, Kleshchonok2018, Hu2021, Marinescu1998}, enabling applications such tunning materials' properties using external fields~\cite{Castro_prl_2007, Yu_Nano-Letters_2009, Kaxiras_nano-lett_2013} and field-assisted exfoliation of layered nanostructures~\cite{Liang_nano-lett_2009, Sidorov_Nanotechnology_2010}.
In addition to these external influences, the structure of the matter system itself plays a crucial role in the resulting QED effects, since molecular vdW interactions inherently reflect the collective response of many atoms to vacuum fluctuations. These interactions cannot be obtained by summing pairwise contributions~\cite{DiStasio-2014-MBD-Review} because each atom couples not only to the vacuum field but also to the fluctuating dipoles of its neighbors. As more atoms or molecules are added, the pattern of electromagnetic fluctuations changes, which alters both the interactions among particles and the overall coupling of the system to the quantum vacuum field. Many-body effects are known to delocalize the charge density along symmetry axes in extended systems such as macromolecules and nanostructures~\cite{ambrosetti2016wavelike}, leading to substantial enhancements in atomic polarizabilities and, consequently, atom-field couplings. The resulting increase in matter-field coupling amplifies QED effects and makes explicit many-body dispersion essential in any realistic QED description of large molecular assemblies. Since solving the fully coupled atom–field dynamics is generally implausible, practical treatments rely on controlled approximations, numerical many-body methods, and simplified models, such as a QED extension of MBD~\cite{Renne1971RetardedQDO, Loris-arxive2025}.

Although research on intermolecular interactions in the weak-coupling regime has been ongoing for decades, many fundamental aspects remain unknown and debated. For instance, how external static fields can (or cannot) influence intermolecular interactions mediated through the fluctuating vacuum EMF~\cite{Fiscelli2020, Hu2021, Reza_PRR2022, Comment-on-FiscelliPRL2021-Abrantes, Comment-on-FiscelliPRL2021-Fiscelli, Karimpour2022CommentField, Kleshchonok2018} by altering the response properties of atoms and whether these changes have classical or quantum mechanical characteristics? Another crucial question is how retardation affects these interactions at intermediate intermolecular distances, comparable to the atomic time scale, which cannot be simply categorized as near- or far-zone. This question was recently considered by Brambilla et al. using an effective field theories approach for two hydrogen atoms~\cite{Brambilla-PRD2017-vdW-H2-full-range}, where they also demonstrated that there is a correction to the London dispersion at short ranges that scales with the interatomic distance as $R^{-3}$. Although this correction resembles an electrostatic term, it is a quantum mechanical correction arising from loop diagrams. Such corrections suggest that the usual truncation of interactions at the leading dipole-dipole order may not accurately represent the physical picture of intermolecular interactions, even for a simple dimer, and particularly when the interactions are mediated through the fluctuating EMF~\cite{Loris-arxive2025}. These examples, along with many others, indicate that we have only scratched the surface of our understanding of intermolecular interactions from the first principles of QED.

\subsection{Strong Matter-Field Coupling}
\begin{figure}[h!]
    \centering\hrulefill\par
    \begin{subfigure}[c]{\textwidth}
        \centering
        \includegraphics[width=0.95\linewidth]{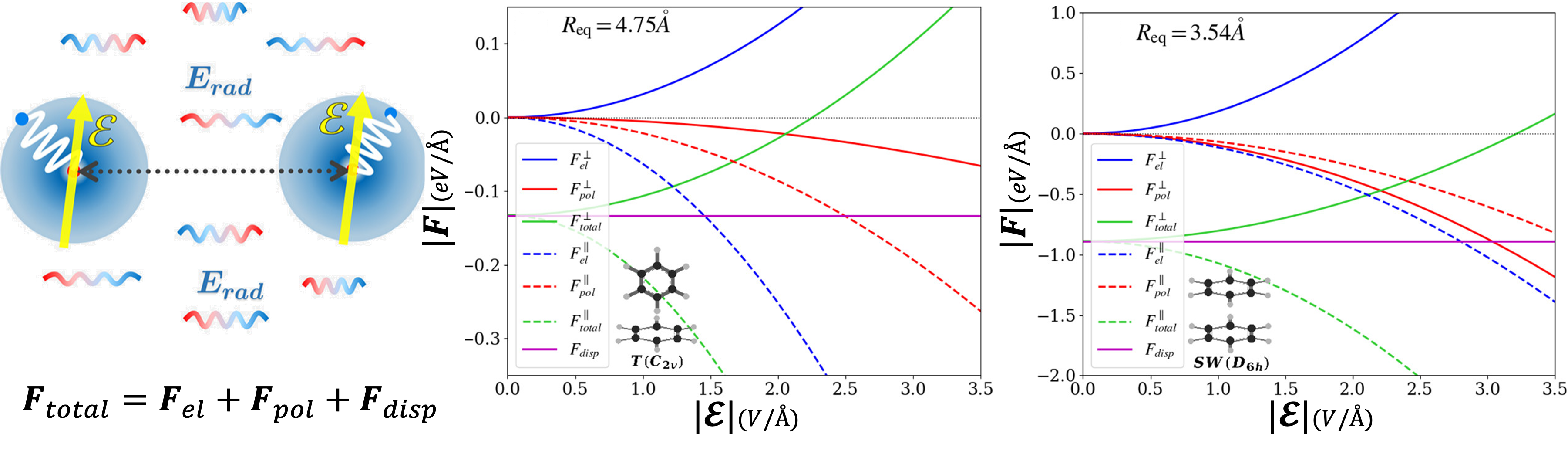}
        \vspace{-0.2cm}
        \caption{\vspace{-0.3cm}}
        \hrulefill\par
        \label{fig:external-field}
    \end{subfigure}
    \begin{subfigure}[c]{\textwidth}
        \centering
        \includegraphics[width=0.95\linewidth]{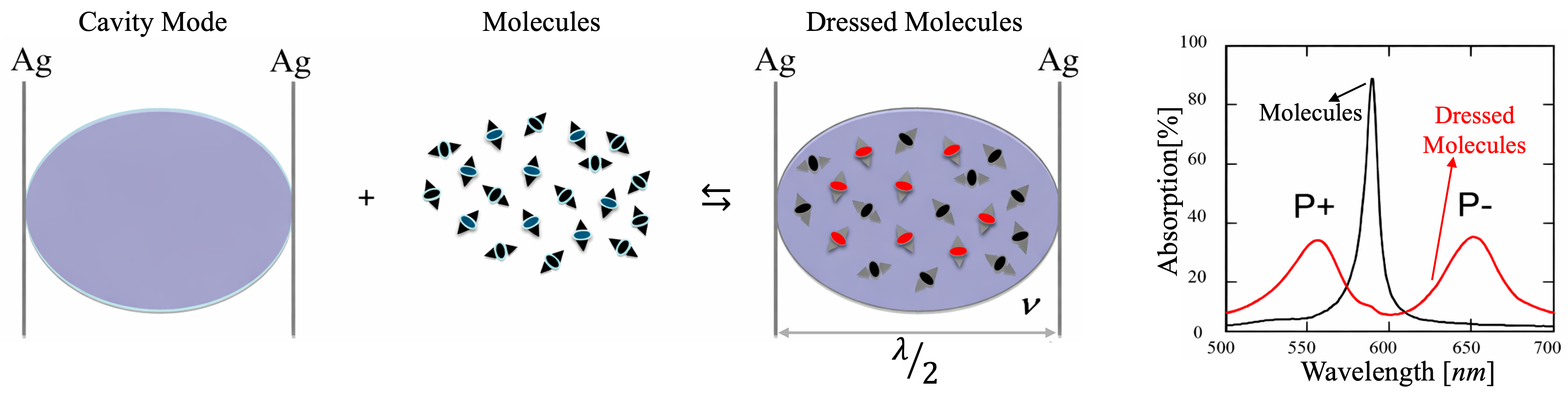}
        \vspace{-0.3cm}
        \caption{\vspace{-0.3cm}}
        \hrulefill\par
        \label{fig:strong-hybrid}
    \end{subfigure}
    \begin{subfigure}[c]{\textwidth}
        \centering
        \includegraphics[width=0.95\linewidth]{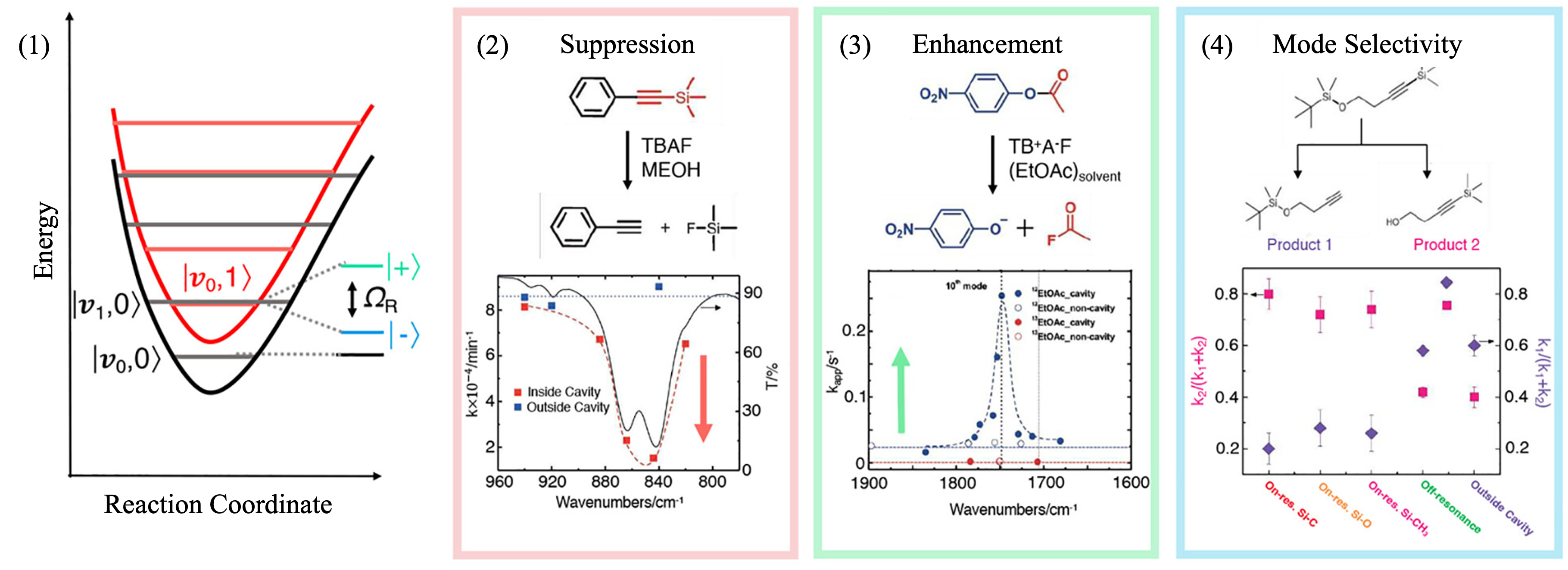}
        \vspace{-0.3cm}
        \caption{\vspace{-0.3cm}}
        \hrulefill\par
        \label{fig:strong-reaction}
    \end{subfigure}
    \caption{\footnotesize {\bf Atoms and Molecules Coupled with Quantum Electromagnetic Field (EMF)}: 
    \textbf{(a)} Interplay between molecular forces induced by an external static field and dispersion force due to the vacuum EMF in two configurations of a benzene dimer demonstrates the possibility of tailoring molecular forces through external fields. Many-body characteristics of dispersion and field-induced polarization forces (taken into account using the MBD method~\cite{DiStasio-2014-MBD-Review}) result in an intricate interplay between these forces and the field-induced electrostatic force. Adapted from~\cite{Reza_JPCL2022}. 
    \textbf{(b)} Electronic strong-coupling of molecules with the cavity EMF yields the matter-field hybridization and formation of upper and lower polaritons that are spectroscopically observable. Adapted from~\cite{Ebbesen2016-Perspective-Hybrid-Light-Matter}.
    \textbf{(c)} Vibrational strong-coupling of molecules with the cavity EMF: (1) Formation of vibrational polariton states $|\pm\rangle$ from hybridization of molecular and cavity excitations. (2) Suppression of reaction rates as a function of cavity photon frequency inside and outside the cavity, with the IR spectrum of the uncoupled molecule. (3) Enhancement of reaction rates under vibrational strong coupling. (4) Cavity-induced mode selectivity in a reaction with two competing products. Adapted from~\cite{Huo2023-Review-PolaritonChemistry}.}
\end{figure}

In contrast to the weak-coupling regime, where matter-field interactions can be treated perturbatively, strong matter-field couplings render QED effects highly influential and beyond the validity of perturbation theory. In the strong-coupling regime, matter and the quantized field must be treated as a hybrid system, as explored in polaritonic chemistry. This approach aims to understand how strong interactions between matter and the quantized field within optical or plasmonic cavities or circuit QED devices modify molecular structure, dynamics, and reactivity~\cite{Ebbesen2012-Photochemical-Reactivity, Ebbesen-ChemPhysChem2013-Polariton-Dynamics, Ebbesen2016-GroundState-Reactivity, Ebbesen2019-Chemical-Reactivity, Flick2022-chemical-reactivity, Feist2017-Polaritonic-Chemistry-Organic-Molecules, Feist2015-Mol-Structure-Bond-Length}, as illustrated, for example, in Figs.~\ref{fig:strong-hybrid} and \ref{fig:strong-reaction}. However, non-perturbative treatments of such hybrid systems are challenging and often intractable, even for small molecules. Therefore, the introduction of simpler, more intuitive model Hamiltonians, along with various approximations, is necessary to simplify these complex matter-field systems. Yet, given the extensive literature on strong matter-field coupling, no single model is universally adopted. Instead, a variety of Hamiltonians, each differing in technical aspects, is employed~\cite{Huo2023-Review-PolaritonChemistry}. 
 
The introduction of various Hamiltonians is primarily driven by the need to balance intuitive understanding with rigorous accuracy across different regimes of light-matter coupling~\cite{Huo2023-Review-PolaritonChemistry, Rubio-ChemRev2023-polaritonic-from-abinitio, Flick-NanoPhotonic2018-Strong-Rev}. Simple models like the Jaynes-Cummings (JC) and Tavis-Cummings (TC) Hamiltonians are frequently employed because they offer analytic solutions and provide clear physical insights, such as the scaling of Rabi splitting with the number of molecules ($\sqrt{N}$)~\cite{Feist2017-Polaritonic-Chemistry-Organic-Molecules, Flick-NanoPhotonic2018-Strong-Rev}. However, these models rely on approximations, specifically the rotating wave approximation and the neglect of dipole self-energy, which become inadequate in the ultrastrong coupling regime or when describing realistic molecular potential energy surfaces~\cite{Ebbesn-PRL2016-Multiple-Rabi, Rubio_PNAS_2021, Rubio-PRX2020-Coupled-Cluster-Polaritons, Flick-NanoPhotonic2018-Strong-Rev, Rubio-PNAS2017-atoms-in-cavities-QEDFT}. Consequently, more rigorous models, such as the Pauli-Fierz (PF) Hamiltonian, are essential to ensure ground-state stability and accurately capture the physics when simpler models fall short~\cite{Huo2023-Review-PolaritonChemistry, Koch2024-Polaritonic-Response-Theory}. In addition, distinct Hamiltonian forms are required to bridge the gap between detailed ab initio simulations of individual molecules and the collective mechanics of macroscopic ensembles observed in experimental cavities\cite{Rubio-JCP2022-Perspective-on-abinitio-polaritonic, Thygesen-NatComm2021-QEDFT-macroscopic-QED, Rubio-PRB2025-full-minimal-coupling-TDDFT, Feist2017-Polaritonic-Chemistry-Organic-Molecules, Rubio-ChemRev2023-polaritonic-from-abinitio}.

Furthermore, the variety of Hamiltonians arises from the mathematical freedom to choose a gauge in QED and the practical constraints of computational simulations. While the Coulomb gauge ($\widehat{\bm{p}}\cdot \widehat{\bm{A}}$) and Dipole gauge ($\widehat{\bm{\mu}}\cdot\widehat{\bm{E}}$) are theoretically equivalent in a complete basis, they yield conflicting numerical results when the electronic basis is truncated, a problem known as ``gauge ambiguity”~\cite{Huo2023-Review-PolaritonChemistry}. This necessitates the development of properly truncated Hamiltonians to ensure physical consistency. Additionally, the choice of Hamiltonian often depends on the specific computational strategy employed. For instance, parameterized QED (pQED) employs precomputed molecular states/potentials as fixed input parameters to enhance computational efficiency, while self-consistent QED (scQED) integrates the photon field into the electronic-structure calculation, necessitating a Hamiltonian that permits variational relaxation of molecular orbitals~\cite{flick2018ab, Huo2023-Review-PolaritonChemistry, Rubio-ChemRev2023-polaritonic-from-abinitio, Rubio-PRX2020-Coupled-Cluster-Polaritons, Flick_PRL_variational_QED}. 

As in the case of weak-coupling, the starting point for deriving QED from first principles is the minimal-coupling Hamiltonian in Coulomb gauge. The interactions of charged particles in the matter system with the quantized electromagnetic field, as described by the principle of minimal coupling of electromagnetism, are incorporated into the system’s kinetic energy. Since the particle’s momentum operator operates nonlocally in the energy space and couples states that are far apart in energy, solutions obtained from this model are often more difficult to converge numerically. In contrast, the multipolar-coupling scheme of molecular QED, derived from the minimal-coupling Hamiltonian through the Power-Zienau-Woolley (PZW) unitary transformation, expresses matter-field interactions in terms of the couplings of matter dipole (or generally multipole) moments to the transverse electric field~\cite{Koch2024-Polaritonic-Response-Theory, Rubio_PNAS_2021, Flick-NanoPhotonic2018-Strong-Rev, Huo2023-Review-PolaritonChemistry}. Since the dipole moment is local in energy space, this gauge is numerically more stable and accurate for truncated basis sets compared to the Coulomb gauge.

Strong coupling between molecular electronic transitions and the quantized electromagnetic field gives rise to hybrid states known as upper and lower polaritons, separated by the Rabi splitting energy, as shown in fig.~\ref{fig:strong-hybrid}. This hybridization creates a distinct spectroscopic splitting in the absorption peak, observable even under vacuum-field conditions without external laser excitation~\cite{Ebbesen2012-Photochemical-Reactivity, Huo2023-Review-PolaritonChemistry}, and is particularly pronounced in organic molecules with large transition dipoles~\cite{Feist2017-Polaritonic-Chemistry-Organic-Molecules, Feist2015-Mol-Structure-Bond-Length}. Fundamentally, this interaction reshapes the system's energy landscape, replacing the Potential Energy Surfaces (PES) of the isolated molecule with distinct Polaritonic Potential Energy Surfaces (PoPES)~\cite{Ebbesen2012-Photochemical-Reactivity}. This reshaping can induce new dynamical features, such as cavity-induced conical intersections, which allow molecules to transition rapidly between energy states. These intersections can guide molecular geometry toward specific products or, through the accumulation of Berry phase shifts, generate destructive interference that effectively blocks specific reaction paths~\cite{Ebbesen2012-Photochemical-Reactivity}. Alternatively, because the lower polariton lies energetically below the uncoupled excited state, it can act as a trap that favors relaxation back to the ground state rather than the forward reaction. This energetic stabilization can suppress the reaction rate, thereby increasing the stability (yield) of the starting reactant~\cite{Ebbesen2012-Photochemical-Reactivity, Huo2023-Review-PolaritonChemistry, Feist2017-Polaritonic-Chemistry-Organic-Molecules, Ebbesen-ChemPhysChem2013-Polariton-Dynamics}. Therefore, electronic strong coupling can fundamentally alter reaction kinetics, pathways, and yields. Hence, in theory, by tuning the cavity frequency, the potential energy surfaces can be engineered to favor one product over another, leading to highly selective isomerization~\cite{Huo2023-Review-PolaritonChemistry}.

The creation of polaritons in a system comprising numerous molecules can occur through collective strong coupling between matter and the quantized field of a cavity. In such scenarios, the hybrid states become delocalized across many molecules, thereby enhancing the material’s transport properties and altering its work function. Consequently, this leads to an increase in the material’s conductivity and has implications for device performance in applications such as organic light-emitting diodes, photovoltaics, and molecular electronics~\cite{Flick_PRL_variational_QED, Ebbesen2016-Perspective-Hybrid-Light-Matter, Ebbesen2012-Photochemical-Reactivity, Huo2023-Review-PolaritonChemistry, Feist2017-Polaritonic-Chemistry-Organic-Molecules}.

In addition to the electronic degrees of freedom of molecules, coupling the cavity mode to molecular vibrational transitions can give rise to new hybrid field-matter states known as vibro-polaritonic states~\cite{Huo2023-Review-PolaritonChemistry, Ebbesen2016-GroundState-Reactivity, Ebbesen2019-Chemical-Reactivity, Simpkins-ChemRev2023-VSC}. Vibrational Strong Coupling (VSC) targets specifically the molecular vibrations (bond stretching and bending) that occur naturally in the electronic ground state. Unlike electronic strong coupling, which requires high-energy photons to excite electrons into excited states, VSC operates in the infrared regime, where vibrational energy levels are accessible via thermal fluctuations. By coupling to these vibrations, the cavity modifies the potential energy landscape that the molecule navigates while in its ground state. The mechanism behind this modification is often described as dynamical caging~\cite{Flick2022-chemical-reactivity, Huo2023-Review-PolaritonChemistry}. The cavity photon mode behaves like an additional solvent environment attached to the reacting bond, thereby increasing friction on the reaction coordinate. This can trap the molecule near the transition-state barrier, suppressing the reaction rate, or modify the flow of vibrational energy between different parts of the molecule, also affecting the reaction rate (see fig.~\ref{fig:strong-reaction}). Since these effects directly influence the vibrational motion, they alter the reaction rate without requiring the molecule to leave its ground electronic state~\cite{Huo2023-Review-PolaritonChemistry, Flick2022-chemical-reactivity, Ebbesen2016-GroundState-Reactivity, Ebbesen2016-Perspective-Hybrid-Light-Matter}.

\subsection{Quantum Electrodynamical DFT}
QEDFT is an exact reformulation of the underlying theory of non-relativistic QED derived from the Pauli-Fierz Hamiltonian in the Coulomb gauge, specifically designed to be computationally feasible for complex systems~\cite{Rubio-PRA2014-QEDFT}. The theoretical foundation of QEDFT lies in a powerful mapping theorem, analogous to the Hohenberg-Kohn and Runge-Gross theorems of standard DFT and time-dependent DFT (TDDFT)~\cite{Tokatly-PRL2013-time-dependent-DFT}. This theorem states that the complete state of a coupled electron-photon system, represented by the full matter-field wavefunction $\Psi({\bmr_j}, {q_\alpha}, t)$, is uniquely determined by a much simpler set of basic variables. These variables include the time-dependent electronic density $n(\bmr,t)$ and the expectation values of the photonic oscillators' coordinates $q_\alpha(t)$, which correspond to the displacement field of each cavity photon mode $\alpha$~\cite{Tokatly-PRL2013-time-dependent-DFT, Rubio-PRA2014-QEDFT, Rubio-PNAS2015-Kohn-Sham-QEDFT, Flick-NanoPhotonic2018-Strong-Rev}. This profound simplification implies that, in principle, the computationally intensive task of computing the complex many-body wavefunctions can be avoided. Instead, all the properties of the system can be derived from these two, significantly lower-dimensional quantities.

To make this mapping practical, QEDFT uses the Kohn-Sham (KS) construction. It replaces the real, interacting system of electrons and photons with a fictitious, non-interacting system that reproduces the same basic variables $n(r,t)$ and $q_\alpha(t)$. This replacement leads to self-consistent Maxwell-Kohn-Sham equations, including a set of single-particle Schrödinger-like equations for the KS electronic orbitals $\phi_i(\bmr,t)$ with an effective potential $v_{\rm KS}(\bmr,t)=v_{\rm ext}(\bmr,t)+v_{\rm H}(\bmr,t)+v_{\rm pxc}(\bmr,t)$ where the new term $v_{\rm pxc}(\bmr,t)$ collects all photon-mediated mean-field plus exchange–correlation effects on electrons, and a set of coupled harmonic oscillator equations for the photon coordinates $q_\alpha(t)$ driven by the electronic dipoles of the matter~\cite{Rubio-PRA2014-QEDFT, Rubio-PNAS2015-Kohn-Sham-QEDFT}. Crucially, these equations are coupled, as the electronic density acts as a source for the photon field, and the photon field acts back on the electrons, creating a self-consistent feedback loop that is absent in standard TDDFT, where the light field is a fixed external input~\cite{Rubio-ChemRev2023-polaritonic-from-abinitio}.

The connection between the real and KS systems lies in the effective KS potential $v_{\rm KS}(\bmr,t)$, which incorporates the conventional terms from electronic DFT (external, Hartree, and electron-electron exchange-correlation) along with the novel photon exchange-correlation (pxc) potential $v_{\rm pxc}$. This term encompasses all non-classical many-body effects arising from electron-photon interactions, including electron self-interaction mediated by emitted photons and other correlations. Consequently, in addition to the conventional density–density response function $\chi_{nn}$, the formalism yields cross-response functions $\chi_{nq}$ and $\chi_{qn}$, as well as the photon–photon response $\chi_{qq}$. This extended response matrix has two significant practical implications. First, electronically excited states manifest as perturbations of the quantized photon field, redefining spectroscopy both conceptually and computationally, from which polaritonic excitations, intrinsic radiative lifetimes, and spectral linewidths emerge naturally without the need for phenomenological reservoirs or damping terms. Second, a direct generalization of the random-phase approximation (RPA), termed polaritonic RPA (pRPA), efficiently captures key cavity-mediated phenomena. Moreover, the inclusion of nonclassical electron–photon exchange–correlation kernels allows the theory to describe multiphoton processes and higher-order light–matter correlations that become important in the strong-coupling regime~\cite{Flick_ACS_photonics_2019, Rubio_PNAS_2021}.

Because the exact form of $v_{\rm pxc}$ is unknown, the primary challenge in applying QEDFT is finding suitable approximations for the photon exchange-correlation functional. A practical approximation scheme developed for QEDFT is based on the Optimized Effective Potential (OEP) method~\cite{Tokatly-PRL2015-OEP-QEDFT, flick2018ab}. The core idea is to derive the potential from an orbital-dependent energy functional. In the lowest order of perturbation theory, the electron-photon exchange energy corresponds to the Lamb shift of the ground state energy. Reformulation of the OEP equations using the Sternheimer equation avoids the need to calculate unoccupied electronic states~\cite{flick2018ab, Rubio-book2012-fundamentals-TDFT}, which is computationally prohibitive, and instead requires only the occupied orbitals. This development enabled the first ab initio QEDFT calculations on real three-dimensional molecules~\cite{flick2018ab}. The accuracy of this exchange-only OEP has been benchmarked against exactly solvable models, demonstrating that it performs well for both weak and strong couplings, except in specific cases of near-degeneracy where multi-photon processes become dominant~\cite{flick2018ab, Tokatly-PRL2015-OEP-QEDFT}.

Although the OEP provides a prominent approximation to the pxc functional, it is not universally applicable and can fail in certain physical situations~\cite{Flick-PRL2025-MBD-QEDFT}. The OEP functional, by construction in its lowest order, primarily accounts for single-photon processes~\cite{flick2018ab, Tokatly-PRL2015-OEP-QEDFT}. Therefore, when multiphoton processes dominate in the strong-coupling regime, and the ground state becomes nearly degenerate with the first excited state, the OEP approximation breaks down because it does not capture higher-order multiphoton processes.
Besides that, even when the QEDFT calculation for the basic variables ($n(\bmr,t)$ and $q_\alpha(t)$) is accurate, calculating other physical observables still requires approximations for those observables' functionals, which is challenging on its own~\cite{Flick_PRL_variational_QED, Rubio-PNAS2017-atoms-in-cavities-QEDFT, Rubio_PNAS_2021}.

Despite its formal elegance, ab initio QEDFT is computationally demanding, much like conventional DFT, and is therefore limited to small systems. This contrasts with most polaritonic chemistry experiments, in which strong coupling is achieved by having many molecules interact simultaneously with a single cavity mode~\cite{Ebbesen2016-Perspective-Hybrid-Light-Matter, Huo2023-Review-PolaritonChemistry}. As a result, QEDFT studies typically simulate only a single molecule and often artificially increase the light–matter coupling to reproduce observable cavity effects~\cite{Flick2022-chemical-reactivity}. These simplifications reduce statistical reliability and leave unresolved the connection between single-molecule simulations and many-molecule experiments. A further limitation is that most QEDFT implementations treat cavity modes as lossless, whereas real cavities exhibit leakage and absorption~\cite{Rubio-PNAS2015-Kohn-Sham-QEDFT, Rubio_PNAS_2021, Flick2022-chemical-reactivity, Flick_natrev_2018, Thygesen-NatComm2021-QEDFT-macroscopic-QED, Rubio-ChemRev2023-polaritonic-from-abinitio}. A practical approach is to combine QEDFT for the electronic subsystem with macroscopic QED for the photonic environment~\cite{Thygesen-NatComm2021-QEDFT-macroscopic-QED, Buhmann-Disp-I-2013}, which incorporates dissipation through Green’s tensors and fluctuation–dissipation relations. Developing efficient hybrid schemes of this kind remains an active area of research.

In conclusion, the mapping theorem at the core of QEDFT generalizes the highly successful principles of DFT to systems where electrons and photons are treated on equal quantum footing. This formally exact framework provides a rigorous foundation for building systematic approximations, distinguishing it from model-based cavity QED approaches. Recent developments, including relativistic and linear-response formulations of QEDFT~\cite{Rubio2025-Dirac-Kohn-Sham, Koch2025-Relativistic-QED}, have extended its reach to realistic molecules and materials, while polaritonic response theory offers new tools for computing optical observables~\cite{Koch2024-Polaritonic-Response-Theory}. Moreover, hybrid schemes that combine density-functional and macroscopic QED formalisms~\cite{Thygesen-NatComm2021-QEDFT-macroscopic-QED} demonstrate how QEDFT can be embedded in lossy or structured photonic environments, thereby bridging the microscopic and continuum scales.

Nevertheless, the practical accuracy of QEDFT still depends critically on the development of reliable electron–photon exchange–correlation functionals and on the validation of predictions under realistic experimental conditions~\cite{Huo2023-Review-PolaritonChemistry, Rubio-JCP2022-Perspective-on-abinitio-polaritonic, Flick-PRL2025-MBD-QEDFT}. Cavity-induced ground-state modifications may be overstated if losses, multimode fields, and disorder are not properly accounted for, and collective or thermal effects can often mimic purely quantum-vacuum signatures~\cite{Rubio-JCP2022-Perspective-on-abinitio-polaritonic, Rubio-ChemRev2023-polaritonic-from-abinitio}. These considerations underscore the necessity for cross-validation with complementary ab initio frameworks, such as QED-coupled-cluster or hybrid DFT+macroscopic-QED approaches, to ensure that QEDFT achieves both conceptual rigor and quantitative reliability in describing cavity-modified molecular and material properties.

\section{Precision Spectroscopy}
\label{sec:Precison_Spec}
Precision physics and spectroscopy enable the measurement of atomic and molecular properties with exceptional accuracy, often reaching 10 to 15 decimal places. The primary objective is to use atoms and molecules as microscopic laboratories to test fundamental physical laws~\cite{Matyus2024-Perspective-Precision-Physics}. For instance, precision measurements of the hydrogen $1S–2S$ transition have tested Quantum Electrodynamics (QED) to 15 decimal places~\cite{Udem-PRL2013-Hexp}. Such experiments advance scientific understanding by verifying the Standard Model and QED, refining fundamental constants, and providing opportunities to identify new physics when discrepancies arise between theoretical predictions and experimental results~\cite{Clark-RevModPhys2018-new-physics-precision, Silkowski2023-QED-Effects-H2, Matyus-PRL2020, Matyus2024-Perspective-Precision-Physics, Matyus2023-Pre-Born-Openheimer-2Body}. To achieve theoretical results that match the experimental precision, one must employ a framework that incrementally constructs the system's energy. Such a framework can be based on non-relativistic QED (nrQED) or relativistic QED (rQED)~\cite{Matyus-ACS2023-Bethe-Salpete-Eq}. The construction of both non-relativistic and relativistic QFTs is briefly discussed in the appendices~\ref{sec:Appendix_A} and \ref{sec:Appendix_Dirac}. Interested readers are referred to consult the references provided in these appendices.

The starting point is to calculate a base energy for the atom or molecule. In the standard Born-Oppenheimer approach, nuclei are assumed to be stationary, and the Schrödinger equation is solved for the electrons. In contrast, the pre-Born-Oppenheimer approach~\cite{Matyus2023-Pre-Born-Openheimer-2Body, Matyus2023-Pre-Born-Openheimer-leading} treats all particles as quantum entities and solves the equation without fixing the nuclear positions. Although this method is computationally intensive, it eliminates errors arising from the assumption of stationary nuclei, yielding more accurate base energies and equations. Then, to obtain the true energy, theoretical models incorporate small corrections to the base equations. These corrections are typically expressed as a series expansion in terms of the fine-structure constant, which quantifies the strength of the electromagnetic interaction.

The framework incorporates several effects, including non-adiabatic, relativistic, and QED corrections~\cite{Matyus-PRL2020}. Non-adiabatic corrections account for nuclear motion and its interaction with electrons, thereby relaxing the assumption of stationary nuclei in the Born-Oppenheimer approach. Relativistic corrections account for mass-velocity effects and electron spin, which become significant as electrons approach the speed of light~\cite{Matyus-PRL2020, Matyus2024-Perspective-Precision-Physics}. QED corrections, which are the most complex to compute, address quantum field theory phenomena such as self-energy, vacuum polarization, and retardation~\cite{Matyus-ACS2023-Bethe-Salpete-Eq, Matyus2023-Pre-Born-Openheimer-leading, Matyus2024-Perspective-Precision-Physics}. When starting from a non-relativistic equation and adding large corrections, these adjustments can be inefficient and susceptible to mathematical divergence. Consequently, studies recommend beginning with a relativistic wave equation, such as the Bethe-Salpeter or Salpeter-Sucher equation, and employing the Dirac equation generalized for systems with two or more particles~\cite{Matyus2023-Pre-Born-Openheimer-2Body, Matyus-ACS2023-Bethe-Salpete-Eq, Matyus2024-Perspective-Precision-Physics}. This strategy incorporates relativity as an intrinsic aspect of the system, thereby reducing computational demands and potentially improving accuracy for heavy atoms where relativistic effects are pronounced.

The Bethe–Salpeter (BS) equationserves as the fundamental field-theoretic starting point for describing a bound state as a pair of particles undergoing an infinite number of scattering events over an infinite lifetime. Because the full BS equation is challenging to solve for atoms and molecules, it is transformed into an exact equal-time variant by integrating over the relative energy. This process yields the Salpeter–Sucher exact equal-time equation, which is formally exact and depends only on the particles' spatial coordinates, making it compatible with traditional computational techniques. However, this equation remains non-linear in energy due to terms that account for retardation and radiative effects. To obtain a practical wave equation, the non-linear Salpeter–Sucher equation is partitioned. By neglecting the non-linear QED terms, the no-pair Dirac–Coulomb–Breit (DCB) equation emerges as the dominant, linear, and Hermitian part.  The framework first solves the no-pair DCB equation, accounting for special relativity and the dominant instantaneous interactions (Coulomb and Breit forces) to all orders in the fine-structure constant. Effects such as retardation, corrections due to the creation of virtual electron-positron pairs, and radiative corrections (self-energy) are subsequently added as perturbations to this relativistic reference. 

A primary contribution of precision spectroscopy is the ongoing validation of QED as the dominant theory for matter–electromagnetic field interactions. Precision spectroscopy of light systems such as hydrogen (H) and helium (He) has provided stringent tests of QED~\cite{Matyus2024-Perspective-Precision-Physics}. Non-Relativistic QED has been the "golden standard" for light elements, yielding high-precision theoretical values that closely match experimental transitions. In systems such as the HD molecule, precision measurements have achieved sub-MHz uncertainty, revealing discrepancies of 1.4 to 1.9~$\sigma$ (\textit{std}) between calculation and experiment~\cite{Silkowski2023-QED-Effects-H2}. These discrepancies indicate that current non-adiabatic and higher-order QED effects are not yet fully understood.

Precision physics has also revealed a significant discrepancy in measuring the proton's size, known as the \textit{proton-size puzzle}.~\cite{RevModPhys2022-proton-size, Guena-PRL2018-proton-size, Udem-Science2017-proton-size} By comparing transition frequencies in electronic hydrogen and muonic hydrogen, it was found that the derived proton charge radius differs significantly depending on whether an electron or a muon orbits the nucleus. These experiments are critical for refining the Rydberg constant and determining accurate nuclear charge radii.

Spectroscopy is increasingly employed to define physical units and constants, rather than just to observe them. Precision measurements of the ${}^{4}He_2^+$ molecular ion are used to determine the polarizability of the helium atom, which is essential for a proposed new pressure standard based on the number density of helium gas atoms~\cite{Matyus-PRL2020, PRL2020-precision-He}. Additionally, experiments with muonium (Mu)~\cite{PRD2019-precision-muonium} have provided insights into the muon’s anomalous magnetic moment, a key area where potential new physics beyond the Standard Model may exist.

Precision measurements of positronium (Ps)~\cite{PRD2019-precision-muonium}, an atom composed of an electron and its antiparticle, are used to test QED without the complexities introduced by nuclear structure. Ps is also a candidate for testing the equivalence principle of gravity on antimatter. For elements with atomic numbers greater than 50, precision mass spectrometry has enabled the investigation of electron binding energies through energy-mass equivalence, thereby testing relativistic QED in high-field regimes~\cite{Matyus2024-Perspective-Precision-Physics}.

Molecular rovibronic mapping, which characterizes the interplay of rotational, vibrational, and electronic energy levels, is fundamental to precision physics because it enables stringent tests of the Standard Model. Measuring these transitions with extreme accuracy allows for direct comparison between experimental data and QED-based calculations. The complex energy landscapes of small molecules have been mapped to unprecedented precision. For instance, in the hydrogen molecule, rovibronic states have been measured and calculated with parts-per-billion accuracy~\cite{Matyus2023-Pre-Born-Openheimer-leading}. For helium ion rotations, the lowest-energy rotational interval of ${}^{4}He_2^+$ was precisely determined to be $70.937589\, cm^{-1}$, establishing a new benchmark for three-electron molecular systems~\cite{Matyus-PRL2020}.

\section{Summary}
The application of quantum field theory (QFT) techniques to chemical systems is an emerging field of research. Hence, one expects to have only scratched the surface of the plethora of effects and approaches that can aid our understanding of the complex behavior of many-body matter/field chemical systems. QFT methods are poised to substantially enhance the arsenal of quantum-chemical techniques while also making the vacuum fields stand out as dynamical quantum entities. QFT may also help to interpret and visualize complex many-body states as fields in real space, and yield novel scaling laws for quantum properties of atoms, molecules, and materials. Connections to the broad and rapidly developing research fields of quantum information theory and quantum computing are also enabled by QFT-inspired tools and language. We thus encourage developers and practitioners of quantum chemistry to deep dive into QFT-inspired methodologies and embrace the QFT-driven path for further development of chemical theory.

\section*{Appendices}
\appendix
\section{A Short QFT Primer}
\label{sec:Appendix_A}

\subsection{General Aspects of QFT}

At its core, QFT treats each field as a quantized entity whose mode amplitudes—when expanded over a complete set of square-integrable, normalized basis functions (the single-particle modes)—become operators acting on a Hilbert space. The spectrum of the corresponding number operators is discrete and labeled by natural numbers, which define the occupation number of a given mode, i.e., they ``count'' the number of particles in that state. This procedure, commonly known as \emph{second quantization}, provides a bridge between classical field amplitudes and the discrete quantum description of matter and radiation. For completeness, we briefly summarize the essential formalism here, referring the reader to these references~\cite{umezawa1982thermo, blasone2011quantum, duncan2012conceptual}.

As in any quantum theory, quantum field theory is based on the specification of a triad:
\begin{itemize}
    \item An algebra of operators $\mathfrak{A}$, within which a subset of Hermitian operators is identified as the \textit{observables};
    \item A Hilbert space $\mathcal{H}$ on which the operators in $\mathfrak{A}$ act;
    \item A Hermitian Hamiltonian operator $\hat{H}\in \mathfrak{A}$ that generates the time evolution of observables, i.e.,
    \begin{equation}
        i\hbar \dfrac{\mathrm{d}\widehat{A}}{\mathrm{d}t} = [\widehat{A},\widehat{H}]\,.
    \end{equation}
\end{itemize}

The fields subjected to the second-quantization procedure include the wavefunctions associated with the fundamental matter degrees of freedom and the vector potential of the electromagnetic field (EMF).

The core of field quantization consists in defining an algebra of creation and annihilation operators $\widehat{a}(\bm{s})$ and $\widehat{a}^{\dagger}(\bm{s})$ for a set of modes, typically identified with a complete infinite orthonormal set of single-particle wavefunctions labeled by $\bm{s}$. The bosonic (fermionic) nature of these operators is encoded in the canonical (anti)commutation relations
\begin{align}
\label{eq:creann_QFT}
[\widehat{a}_{\bm{s}},\widehat{a}^{\dagger}_{\bm{s}'}]_{\pm}
&=\delta_{\bm{s}\bm{s}'} \,,
\qquad
[\widehat{a}_{\bm{s}},\widehat{a}_{\bm{s}'}]_{\pm}
=
[\widehat{a}^{\dagger}_{\bm{s}},\widehat{a}^{\dagger}_{\bm{s}'}]_{\pm}=0\,,
\end{align}
where $[\widehat{A},\widehat{B}]_{\pm}=\widehat{A}\widehat{B}\pm\widehat{B}\widehat{A}$, with the $(-)$ sign for bosons and the $(+)$ sign for fermions. These operators represent the quantized mode amplitudes of a quantum field. The operators
\[
\widehat{n}_{\boldsymbol{s}} = \widehat{a}_{\boldsymbol{s}}^{\dagger}\widehat{a}_{\boldsymbol{s}}
\]
are Hermitian number operators whose spectra consist of the natural numbers; they measure the number of particles occupying the mode labeled by $\boldsymbol{s}$. These operators act on a vector space spanned by Fock states of the form
\[
|n_{\boldsymbol{s}_1}, n_{\boldsymbol{s}_2}, \ldots, n_{\boldsymbol{s}_i}, \ldots \rangle ,
\]
where the non-negative integer $n_{\boldsymbol{s}_i}$ denotes the number of particles in the single-particle state $\boldsymbol{s}_i$. This space includes the \emph{vacuum state}
\[
|\bm{0}\rangle = |0,0,0,\ldots\rangle,
\]
which contains no particles in any mode. The creation and annihilation operators introduced in Eq.~\eqref{eq:creann_QFT} act on these basis states according to
\begin{align}
\label{eq:creation_annihilation}
\widehat{a}_{\boldsymbol{s}_i}
|n_{\boldsymbol{s}_1}, \ldots, n_{\boldsymbol{s}_i}, \ldots \rangle
&=
\sqrt{n_{\boldsymbol{s}_i}}
|n_{\boldsymbol{s}_1}, \ldots, n_{\boldsymbol{s}_i}-1, \ldots \rangle ,\\
\widehat{a}^{\dagger}_{\boldsymbol{s}_i}
|n_{\boldsymbol{s}_1}, \ldots, n_{\boldsymbol{s}_i}, \ldots \rangle
&= 
\sqrt{n_{\boldsymbol{s}_i}+1}
|n_{\boldsymbol{s}_1}, \ldots, n_{\boldsymbol{s}_i}+1, \ldots \rangle .
\end{align}
For fermionic creation and annihilation operators, the occupation numbers are restricted to $n_{\boldsymbol{s}} \in \{0,1\}$, since the Pauli exclusion principle implies $(\widehat{a}_{\boldsymbol{s}}^{\dagger})^2 = 0\,$.

\paragraph{Construction of the Fock Space in QFT}

\paragraph{Representations of the Canonical (Anti)commutation Relations}

To define a quantum theory, one must specify a Hilbert space carrying a representation of the algebra of creation and annihilation operators. A crucial distinction arises between ordinary quantum mechanics, where a \textit{finite} number of modes is involved, and quantum field theory (or many-body quantum mechanics) in the thermodynamic or continuum limit, where an \textit{infinite} number of modes must be considered.

In the finite-mode case, which is typically encountered in practical quantum-chemical applications, the Fock space is constructed from occupation-number states of the form
\begin{equation}
\lvert n_1, n_2, \ldots \rangle .
\end{equation}
Here the indices label a countable set of single-particle modes, such as atomic or molecular orbitals. The resulting Fock space is separable, and different choices of orthonormal single-particle bases correspond to different but unitarily equivalent representations of the same physical Hilbert space. This situation closely parallels the familiar freedom in quantum chemistry to choose different orbital bases, which provide alternative but equivalent descriptions of the same many-electron state.

A qualitatively different situation arises when infinitely many degrees of freedom are involved. In this case, the canonical (anti)commutation relations admit \emph{unitarily inequivalent representations}, which cannot be related to one another by unitary transformations acting within a single Hilbert space.

A general procedure for constructing a particular representation of the canonical (anti)commutation relations consists of:
\begin{enumerate}
    \item Specifying an algebra of creation and annihilation operators
    \(
    \{ \hat a_s, \hat a_s^\dagger \}
    \)
    associated with an infinite set of modes.
    
    \item Choosing a reference (vacuum) state \(\lvert \mathbf{0} \rangle\) such that
    \begin{equation}
    \hat a_s \lvert \mathbf{0} \rangle = 0 \qquad \forall s \,.
    \end{equation}
    
    \item Constructing the vector space spanned by occupation-number states with finite particle number, for example
    \begin{equation}
    \mathcal{S} =
    \left\{
    \lvert n_1, n_2, \ldots \rangle
    \;\middle|\;
    \sum_i n_i = N,\;
    N \in \mathbb{N}_0
    \right\}.
    \end{equation}
\end{enumerate}

The closure of the vector space generated by \(\mathcal{S}\) defines a separable Hilbert space, usually referred to as the (bosonic or fermionic) Fock space associated with the chosen vacuum. In systems with infinitely many degrees of freedom, different choices of vacuum may lead to representations that are unitarily inequivalent. These inequivalent representations do not reflect different basis-set choices, but rather correspond to physically distinct realizations of the system, such as different vacuum states, broken-symmetry phases, or collective many-body backgrounds. A relevant example in this sense is provided by the Hilbert spaces of the electromagnetic field in the presence of different boundary conditions. This feature underlies the emergence of quasiparticles, phase transitions, and effective field descriptions commonly encountered in condensed-matter physics and QED~\cite{umezawa1982thermo,umezawa1995advanced,blasone2011quantum}.

\paragraph{Construction of Non-Relativistic QFTs}

Within this framework, the non-relativistic field operator $\widehat{\psi}(\boldsymbol{x},t)$ and its Hermitian conjugate $\widehat{\psi}^{\dagger}(\boldsymbol{x},t)$ are defined as
\begin{equation}
\widehat{\boldsymbol{\psi}}(\boldsymbol{x},t)
=
\sum_{\boldsymbol{s}}
\boldsymbol{f}_{\boldsymbol{s}}(\boldsymbol{x})\,
\widehat{a}_{\boldsymbol{s}}(t)\,,
\end{equation}
where $\boldsymbol{f}_{\boldsymbol{s}}(\boldsymbol{x}) = f_{\boldsymbol{s},i}\,\boldsymbol{e}_i$ are mode functions that transform in the same way under spatial rotations and spin transformations as the field operator $\widehat{\boldsymbol{\psi}}(\boldsymbol{x},t)$, i.e., they form a set of orthonormal spin orbitals. The field operators satisfy the equal-time (anti)commutation relations
\begin{align}
[\widehat{\psi}_i(\boldsymbol{x},t), \widehat{\psi}_j^{\dagger}(\boldsymbol{y},t)]_{\pm}
&=
\delta_{ij}\delta^{(3)}(\boldsymbol{x}-\boldsymbol{y}),\\
[\widehat{\psi}_i(\boldsymbol{x},t), \widehat{\psi}_j(\boldsymbol{y},t)]_{\pm}
&=
[\widehat{\psi}_i^{\dagger}(\boldsymbol{x},t), \widehat{\psi}_j^{\dagger}(\boldsymbol{y},t)]_{\pm}=0\,.
\end{align}
Here $[\cdot,\cdot]_{-}$ denotes commutators for bosons, and $[\cdot,\cdot]_{+}$ denotes anticommutators for fermions.

The mapping between the first-quantized, particle-based representation and the second-quantized, field-based representation can be constructed as follows for single-particle and two-particle operators:
\begin{align}
&\widehat{O}^{(I)}(\hat{\bm{p}},\hat{\bm r})
=\sum_{i,j} O^{(I)}_{ij}(\hat{\bm{p}},\hat{\bm r})\, \bm e_i \otimes \bm e_j,
\\
& \widehat{O}^{(II)}(\hat{\bm{p}},\hat{\bm r}; \hat{\bm{p}}', \hat{\bm r}')
=\sum_{i,j,k,l} O^{(II)}_{ijkl}(\hat{\bm{p}},\hat{\bm r}; \hat{\bm{p}}', \hat{\bm r}')\,
(\bm e_i \otimes \bm e_j)\otimes(\bm e_k \otimes \bm e_l)\,,
\end{align}
respectively. Expanding the field operators in an orthonormal single-particle basis $\{\bm f_{\bm s}\}$ yields
\begin{align}
&\hat{O}^{I}
= \sum_{\bm{s},\bm{s}'} \langle \bm{f}_{\bm s}|\widehat{O}^{(I)}(\hat{\bm{r}},\hat{\bm{p}}) |\bm{f}_{\bm s'}\rangle \,
\widehat{a}^{\dagger}_{\bm s} \widehat{a}_{\bm s'}
\\
\nonumber &=
\sum_{\bm{s},\bm{s}'}\left[\sum_{i,j}\int
f^{*}_{\bm{s},i}(\bm{r})\, O^{(I)}_{ij}(\bm{r},-i\hbar\nabla_{\bm{r}})\, f_{\bm{s}',j}(\bm{r})\,\mathrm{d}^3\bm{r}\right]
\widehat{a}^{\dagger}_{\bm s} \widehat{a}_{\bm s'} ,
\\[4pt]
&\hat{O}^{II}
= \dfrac{1}{2}\sum_{\bm{s},\bm{w},\bm{s}',\bm{w}'}  \langle \bm{f}_{\bm{s}}\bm{f}_{\bm{w}}|\widehat{O}^{(II)}(\hat{\bm{p}},\hat{\bm{r}},\hat{\bm{p}}',\hat{\bm{r}}')| \bm{f}_{\bm{s}'}\bm{f}_{\bm{w}'} \rangle 
\widehat{a}_{\bm{s}}^{\dagger}\widehat{a}_{\bm{w}}^{\dagger} \widehat{a}_{\bm{s}'}\widehat{a}_{\bm{w}'}
\\
\nonumber &=
\dfrac{1}{2}\sum_{\bm{s},\bm{w},\bm{s}',\bm{w}'}  
\left[\sum_{i,j,k,l}\iint\,\,f^{*}_{s,i}(\bm{r})f^{*}_{w,j}(\bm{r}')
O_{ijkl}^{(II)}(-i\hbar \nabla_{\bm{r}}, \bm{r}, -i\hbar \nabla_{\bm{r}'},\bm{r}') f_{s',k}(\bm{r})f_{w',l}(\bm{r}')
\mathrm{d}^3\bm r\,\mathrm{d}^3\bm r'
\right]\\
\nonumber &\times\widehat{a}^{\dagger}_{\bm s}\widehat{a}^{\dagger}_{\bm w}\widehat{a}_{\bm s'}\widehat{a}_{\bm w'} .
\end{align}

This mapping allows one to express any Hamiltonian operator $\hat{H}$ appearing in Schr\"odinger equations in a corresponding second-quantized form. Within this framework, the electronic Coulomb Hamiltonian generally adopted in quantum chemistry can be rewritten in the second-quantization formalism as
\begin{equation}
\widehat{H}_{\rm Coul}=\sum_{sw} h_{sw} \widehat{a}_{s}^{\dagger}\widehat{a}_{w}+\dfrac{1}{2} \sum_{ss'ww'} h_{sws'w'}  \widehat{a}_{s}^{\dagger}\widehat{a}^{\dagger}_{w}
\widehat{a}_{s'}\widehat{a}_{w'}\,,
\end{equation}
where the single-particle and two-particle Hamiltonian contributions read
\begin{align}
    h_{sw}&= \int \psi^{*}_{s}(\bm{r}) \left(-\hbar^2 \dfrac{\nabla_{\bm{r}}^2}{2m_e}-\sum_{A=1}^{N_n} \dfrac{Z_A e^2}{4\pi\varepsilon_0 \|\bm{r}-\bm{R}_A\|} \right)\psi_{w}(\bm{r}) \,\mathrm{d}^3\bm{r},\\
    h_{sws'w'}&= e^2 \iint \dfrac{\psi^{*}_{s}(\bm{r})\psi^{*}_{w}(\bm{r}')\psi_{s'}(\bm{r})\psi_{w'}(\bm{r}')}{4\pi\varepsilon_0 \|\bm{r}-\bm{r}'\|}
   \,\mathrm{d}^3\bm{r}\,\mathrm{d}^3\bm{r}'\,.
\end{align}
where $\{\psi_{s}(\bm{r})\}_{s}$ is a set of orthonormalized orbitals.

\paragraph{Construction of relativistic QFTs}
In relativistic field theories—such as quantum electrodynamics or the Dirac theory of electrons, which is essential in quantum chemistry for describing relativistic electrons—defining fields and Hamiltonian operators that transform covariantly under Lorentz transformations involves significant subtleties. This helps explain the wide range of methods that have been proposed for constructing quantum field theory, including canonical quantization, path-integral quantization, and axiomatic quantum field theory.

A possible approach is to define the field operators for free theories, which transform according to finite-dimensional representations of the Lorentz group—e.g., scalar (spin-0), massive spin-$1/2$ (e.g., electrons), or massless spin-$1$ (e.g., photons)—as well as under discrete symmetry operations (parity, charge conjugation, and time reversal).

Local Hamiltonians consistent with special relativity can then be constructed by adding to the free Hamiltonian an interaction Hamiltonian density expressed as a local polynomial in the fields and their derivatives, while imposing the microcausality condition. Assuming that the Hamiltonian operator $\widehat{H}$ can be written in terms of a local Hamiltonian density $\widehat{h}(\bm{r}, t)$ such that
\begin{equation}
    \widehat{H}(t) = \int_{\mathbb{R}^3} \widehat{h}(\bm{r}, t)\, \mathrm{d}^3\bm{r},
\end{equation}
the Hamiltonian density operators evaluated at spacelike-separated spacetime points $x$ and $y$ must commute:
\begin{equation}
    [\widehat{h}(x), \widehat{h}(y)] = 0 \quad \text{for} \quad (x - y)^2 < 0.
\end{equation}
Relativistic quantum field operators can be constructed starting from plane-wave solutions of relativistic wave equations (e.g., Klein--Gordon or Dirac), i.e.,
\begin{equation}
\label{eq:gen_QFT_plainwaves}
\widehat{\Phi}_\alpha(x)=
\sum_{\sigma}
\int \frac{\mathrm{d}^3\bm{p}}{(2\pi)^3\,2E_{\bm{p}}}
\left[
u_\alpha(\bm{p},\sigma)\,\widehat{a}(\bm{p},\sigma)\,
e^{-\frac{i}{\hbar}p\cdot x}
+
\overline{v}_\alpha(\bm{p},\sigma)\,\widehat{b}^\dagger(\bm{p},\sigma)\,
e^{\frac{i}{\hbar}p\cdot x}
\right],
\end{equation}
where
\[
p^\mu=\left(\frac{E_{\bm{p}}}{c},\,\bm{p}\right),
\qquad
E_{\bm{p}}=\sqrt{\bm{p}^{\,2}c^2+m^2c^4}.
\]
Here, $\alpha$ labels the components of a finite-dimensional Lorentz representation, $\sigma$ labels spin/helicity states, and the functions $u_\alpha(\bm{p},\sigma)$ and $v_\alpha(\bm{p},\sigma)$ correspond to the components of covariant amplitudes. Such amplitudes are built from irreducible finite-dimensional representations of the stabilizer (little) subgroup of the Lorentz group that leaves a given momentum $p^\mu$ invariant. This subgroup differs for massive and massless particles:
\begin{itemize}
    \item For \textbf{massive particles}, the little group is isomorphic to $SO(3)$, the rotation group in the rest frame. States are labeled by mass $m>0$ and spin $s = 0, \frac{1}{2}, 1, \dots$, with $2s+1$ spin states.

    \item For \textbf{massless particles}, the little group is isomorphic to the Euclidean group $ISO(2)$, and its finite-dimensional unitary representations are labeled by the \textit{helicity} $\lambda = \pm s$ (for particles with definite spin $s$) and the momentum magnitude $|\bm{p}|$ (or, equivalently, the energy $E = |\bm{p}|c$). Helicity is the projection of spin along the direction of motion and is Lorentz-invariant for massless particles.
\end{itemize}

Note that, unlike in non-relativistic field theories, the relativistic field operator is a linear combination of creation and annihilation operators. This is a direct consequence of microcausality, which requires that field operators at spacelike separation commute (for bosons) or anticommute (for fermions), i.e.,
\begin{equation}
    [\hat{\Phi}_{\alpha}(x),\hat{\Phi}^{\dagger}_{\beta}(y)]_{\pm} = 0 \qquad (x-y)^2 <0\,.
\end{equation}
This structure also reflects the fact that each relativistic field generally describes two types of excitations: particles and antiparticles. The operator $\widehat{a}(\bm{p},\sigma)$ annihilates a particle with momentum $\bm{p}$ and spin $\sigma$, while $\widehat{b}^\dagger(\bm{p},\sigma)$ creates an antiparticle with the same quantum numbers. Their Hermitian conjugates, $\widehat{a}^\dagger(\bm{p},\sigma)$ and $\widehat{b}(\bm{p},\sigma)$, create particles and annihilate antiparticles, respectively. For neutral fields—i.e., fields that are their own antiparticles—the particle and antiparticle states coincide. In such cases, the field operator contains only one set of creation and annihilation operators. For example, Majorana fermions satisfy $\widehat{a}^{(\dagger)}=\widehat{b}^{(\dagger)}$, while the photon field (a real vector field) has no distinct antiparticle operator and is expanded solely in terms of $\widehat{a}$ and $\widehat{a}^{\dagger}$.

The (anti)commutation relations of the field operators follow from the canonical (anti)commutation relations imposed on the creation and annihilation operators, which for free fields read
\begin{align}
[\widehat{a}(\bm{p},\sigma), \widehat{a}^\dagger(\bm{p}',\sigma')]_{\pm}
&=
[\widehat{b}(\bm{p},\sigma), \widehat{b}^\dagger(\bm{p}',\sigma')]_{\pm}
=
(2\pi)^3\,2E_{\bm{p}}\,
\delta^{(3)}(\bm{p}-\bm{p}')\,\delta_{\sigma\sigma'}\,,
\end{align}
with all other (anti)commutators vanishing. The upper (lower) sign refers to fermionic (bosonic) fields, in accordance with the spin--statistics theorem.

\paragraph{Interacting fields and physical quasi-particle fields}
Finally, it is worth emphasizing that in quantum field theory there exists a \emph{weak operator identity}, denoted by $\overset{W}{=}$, which holds only at the level of matrix elements between states of a complete basis. This weak identity relates two sets of field operators acting on Fock space: the interacting Heisenberg fields $\{\widehat{\psi}_A\}_{A=1,\dots,N}$ and the physical quasi-particle fields $\{\widehat{\phi}_a\}_{a=1,\dots,N}$. Explicitly, one may write
\begin{align}
\widehat{\psi}_{A}(x)\overset{W}{=}\;&
\rho_{A}(x)\,\widehat{\mathbb{I}}
+\sum_{a} Z_{Aa}^{1/2}\,\widehat{\phi}_a(x)
+\sum_{ab}\int \mathrm{d}^4 y_1\,\mathrm{d}^4 y_2\,
K_{Aab}(x;y_1,y_2)\,
:\widehat{\phi}_a(y_1)\widehat{\phi}_b(y_2): \nonumber\\
&+\sum_{abc}\int \mathrm{d}^4 y_1\,\mathrm{d}^4 y_2\,\mathrm{d}^4 y_3\,
K_{Aabc}(x;y_1,y_2,y_3)\,
:\widehat{\phi}_a(y_1)\widehat{\phi}_b(y_2)\widehat{\phi}_c(y_3):
+\cdots .
\end{align}
Here $\rho_A(x)$ is a c-number function, which can be non-vanishing only for bosonic fields, $Z_{Aa}^{1/2}$ are field-strength renormalization constants, and $K_{Aab}(x;y_1,y_2)$, $K_{Aabc}(x;y_1,y_2,y_3)$, $\ldots$ are (generally nonlocal) convolution kernels. Normal ordering is defined with respect to the vacuum of the asymptotic quasi-particle fields. The kernels may be chosen such that, when evaluated between physical quasi-particle states, the interacting Hamiltonian is weakly equivalent to an effective free Hamiltonian expressed in terms of the creation and annihilation operators $\hat{a}_s^\dagger,\hat{a}_s$ associated with the quasi-particle fields, namely
\begin{equation}
\widehat{H}[\{\widehat{\psi}_A\}_A]\overset{W}{=}
\sum_{s} h_{s}\left(\hat{a}_s^{\dagger}\hat{a}_{s}
\pm \frac{1}{2}\right)=\widehat{H}_0[\{\widehat{\phi_a}\}_a]\,,
\end{equation}
where the plus sign applies to bosonic modes and the minus sign to fermionic modes. In quantum chemistry and condensed-matter applications, this weak equivalence underlies the practical success of quasiparticle descriptions, in which interacting many-electron systems are effectively described in terms of dressed single-particle excitations, despite the absence of exact integrability. Such a conceptual picture provides a formal justification for the widespread use of effective single-particle descriptions, normal ordering with respect to a reference determinant, and perturbative or coupled-cluster expansions in electronic-structure theory. In Green's-function formulations, the same structure appears through the emergence of quasiparticle poles of the one-body propagator, corresponding to dressed electronic excitations with renormalized energies and spectral weights.

For applications in quantum chemistry, the relevant relativistic fields are the fermionic relativistic field—representing relativistic electrons, particularly core electrons in heavy elements—and the electromagnetic field.

\section{Dirac Free Equation}
\label{sec:Appendix_Dirac}
The free field describing relativistic fermions is the Dirac field $\widehat{\psi}(x)$, a four-component complex spinor field satisfying the Dirac equation
\begin{equation}
\label{eq:Dirac-equation}
i\hbar\,\gamma^{\mu}\partial_{\mu}\psi(x)-mc\,\psi(x)=0\,,
\end{equation}
where $x^{0}=ct$ is the time coordinate, $x^{i}$ ($i=1,2,3$) are spatial coordinates, and $\gamma^{\mu}$ are the $4\times4$ Dirac matrices in the Dirac representation, satisfying the Clifford algebra relation $\{\gamma^{\mu}, \gamma^{\nu}\}=2\eta^{\mu \nu}\mathbb{I}_{4\times 4}$, i.e.,
\begin{equation}
\gamma^{0}=\begin{pmatrix}
\mathbb{I}_2 & 0 \\
0 & -\mathbb{I}_2
\end{pmatrix},
\qquad
\gamma^{i}=\begin{pmatrix}
0 & \sigma_i \\
-\sigma_i & 0
\end{pmatrix},
\end{equation}
with $\mathbb{I}_2$ the $2\times2$ identity matrix and $\sigma_i$ the Pauli matrices
\begin{equation}
\sigma_{1}=\begin{pmatrix}0 & 1\\ 1 & 0\end{pmatrix},
\quad
\sigma_{2}=\begin{pmatrix}0 & -i\\ i & 0\end{pmatrix},
\quad
\sigma_{3}=\begin{pmatrix}1 & 0\\ 0 & -1\end{pmatrix}.
\end{equation}

The Dirac field admits a mode decomposition of the form in Eq.~\eqref{eq:gen_QFT_plainwaves}, where $u=u(p)$ and $v=v(p)$ are four-component spinors satisfying
\begin{equation}
(\gamma^{\mu}p_{\mu}-mc\,\mathbb{I}_4)\,\bm{u}(p)=0,
\qquad
(\gamma^{\mu}p_{\mu}+mc\,\mathbb{I}_4)\,\bm{v}(p)=0\,.
\end{equation}

Applying minimal coupling to the Dirac Hamiltonian yields the charge density and current density operators
\begin{align}
\widehat{\rho}(\bm r,t)
&= -e\,\widehat{\overline{\psi}}(\bm r,t)\gamma^{0}\widehat{\psi}(\bm r,t),\\
\widehat{\bm J}(\bm r,t)
&= -e\,c\,\widehat{\overline{\psi}}(\bm r,t)\,
\boldsymbol{\gamma}\,\widehat{\psi}(\bm r,t),
\end{align}
where $\widehat{\overline{\psi}}=\widehat{\psi}^{\dagger}\gamma^{0}$ and $\boldsymbol{\gamma}=(\gamma^1,\gamma^2,\gamma^3)$. In many quantum-chemical applications, one works in a no-pair 
approximation, so that the field is projected onto the positive-energy (electron) subspace and 
the charge and current densities can be expressed solely in terms of electron creation and 
annihilation operators $\widehat{a}(\bm{p},s)$ and $\widehat{a}^{\dagger}(\bm{p},s)$, with 
reference to Eq.~\eqref{eq:gen_QFT_plainwaves}. Relativistic electrons in the presence of an 
external electric field generated by atomic nuclei (or a more general electromagnetic field, if 
nuclear total angular momentum is included) are usually described by effective Hamiltonians 
derived from a full relativistic quantum electrodynamics picture. The mathematical treatment 
required to derive the spectrum of these theories goes beyond the scope of the current review, 
and we refer to the specialized literature for further details~\cite{roos2004relativistic,reiher2015relativistic,liu2020essentials}.

\section{Molecular Quantum Electrodynamics in a Nutshell}
\label{sec:Appendix_B}

The multipolar, or Power--Zienau--Woolley (PZW), formalism for nonrelativistic
electrodynamics is obtained through a canonical transformation of the classical
electromagnetic--matter phase space.
Upon quantization, this transformation is implemented by a unitary operator.
The resulting representation is particularly well suited to describing the
interaction between localized, distinguishable charge distributions and the
electromagnetic field in terms of electric and magnetic multipole moments.

The central idea is to formulate the coupling between matter and the
electromagnetic field (EMF) through the polarization $\bm{P}(\bm{r},t)$ and
magnetization $\bm{M}(\bm{r},t)$ fields, which are composite quantities
constructed from the microscopic degrees of freedom of matter, in close analogy
with macroscopic molecular electrodynamics.
From a mathematical perspective, this construction is naturally embedded in the
four-dimensional covariant formulation of Maxwell's equations in terms of the
Faraday tensor
\begin{equation}
F_{\mu\nu}=\partial_\mu A_\nu-\partial_\nu A_\mu\,,
\end{equation}
where $\mu,\nu=0,1,2,3$ label the Cartesian spacetime coordinates $(ct,x,y,z)$, and
indices are raised and lowered using the Minkowski metric
$\eta_{\mu\nu}=\mathrm{diag}(1,-1,-1,-1)$. The four-current is defined as
$J^\mu=(c\rho,\bm{J})$, and Maxwell's equations can be written compactly as
\begin{equation}
\label{eq:MaxwellEoM}
\partial_\mu F^{\mu\nu}=\mu_0 J^\nu,
\qquad
\partial_{[\mu}F_{\nu\rho]}=0\,,
\end{equation}
where square brackets on the indices indicate full antisymmetrization.

Within this framework, it is natural to introduce an antisymmetric rank-two
polarization tensor $P^{\mu\nu}$ such that
\begin{equation}
J^\nu \equiv J^\nu_{\mathrm{free}}+\partial_\mu P^{\mu\nu}\,,
\end{equation}
with the supplementary requirement that the continuity equation be satisfied,
\begin{equation}
\partial_\mu J^\mu
=
\partial_\mu J^\mu_{\mathrm{free}}
+\partial_\mu\partial_\nu P^{\mu\nu}
=0\,.
\end{equation}
Assuming that all charges are bound, so that there are no freely moving charges or
currents, the charge and current associated with freely moving charges vanish,
i.e., $J^\mu_{\mathrm{free}}=0$.
In this case, the antisymmetric tensor
\begin{equation}
\label{eq:Pmunu_def}
P^{0i}=P^{i},\qquad
P^{ij}=\epsilon^{ijk}M_k,\qquad
P^{\mu\nu}=-P^{\nu\mu},
\end{equation}
leads directly to the constitutive relations for the polarization and
magnetization fields,
\begin{align}
\label{eq:constEQ_Pol}
\nabla\cdot \bm{P}(\bm{r},t) &= -\rho(\bm{r},t),\\
\label{eq:constEQ_Mag}
\nabla\times \bm{M}(\bm{r},t) &= \bm{J}(\bm{r},t)-\partial_t\bm{P}(\bm{r},t).
\end{align}

Equations~\eqref{eq:constEQ_Pol} and~\eqref{eq:constEQ_Mag} do not uniquely define
the polarization (and, consequently, the magnetization) field. For bounded
charge distributions, the polarization depends on the arbitrary choice of a
reference point (pole of reduction). For a system of $N_{\rm mol}$ molecules, we
introduce one pole $\bm{O}_\xi$ for each molecule
$\xi=1,\dots,N_{\rm mol}$ and write
\begin{equation}
\bm{P}(\bm{r},t)=\sum_{\xi=1}^{N_{\rm mol}}\bm{P}_\xi(\bm{r},t;\bm{O}_\xi),
\qquad
\bm{M}(\bm{r},t)=\sum_{\xi=1}^{N_{\rm mol}}\bm{M}_\xi(\bm{r},t;\bm{O}_\xi),
\end{equation}
together with $\rho(\bm{r},t)=\sum_\xi \rho_\xi(\bm{r},t)$ and
$\bm{J}(\bm{r},t)=\sum_\xi \bm{J}_\xi(\bm{r},t)$.

In general,
\begin{equation}
\label{eq:P_general}
\bm{P}_{\xi}(\bm{r},t;\bm{O}_{\xi})
=
\int
\bm{\mathcal{G}}(\bm{r},\bm{r}';\bm{O}_{\xi})\,
\rho_{\xi}(\bm{r}',t)\,
\mathrm{d}^3\bm{r}'\,,
\end{equation}
where $\bm{\mathcal{G}}(\bm{r},\bm{r}';\bm{O}_\xi)$ is a vector-valued Green
function satisfying
\[
\nabla_{\bm{r}}\cdot
\bm{\mathcal{G}}(\bm{r},\bm{r}';\bm{O}_{\xi})
=
-\delta^{(3)}(\bm{r}-\bm{r}')\,.
\]
An explicit representation is
\begin{equation}
\label{eq:G_decomposition}
\bm{\mathcal{G}}(\bm{r},\bm{r}';\bm{O}_{\xi})
=
\bm{\mathcal{G}}^{\parallel}(\bm{r},\bm{r}';\bm{O}_{\xi})
+
\int_0^1
(\bm{r}'-\bm{O}_{\xi})\,
\delta^{(3)}\!\bigl[(\bm{O}_{\xi}-\bm{r})
+\lambda(\bm{r}'-\bm{O}_{\xi})\bigr]\,
\mathrm{d}\lambda\,,
\end{equation}
where
\[
\bm{\mathcal{G}}^{\parallel}(\bm{r},\bm{r}';\bm{O}_{\xi})
=
-\nabla_{\bm{r}}
\left[
\frac{1}{4\pi\|\bm{r}-\bm{r}'\|}
\right]
\]
denotes the longitudinal component of the Green function, while the
integral term represents its transverse contribution.

With these conventions, the polarization field may be written as
\begin{align}
\label{eq:P_explicit}
\bm{P}_{\xi}(\bm{r},t;\bm{O}_{\xi})
&=
Q_{\xi}\,\bm{\mathcal{G}}^{\parallel}(\bm{r};\bm{O}_{\xi})
\\
&\quad+
\int\!\!\int_0^1
(\bm{r}'-\bm{O}_{\xi})\,
\delta^{(3)}\!\bigl[\lambda(\bm{r}'-\bm{O}_{\xi})
-(\bm{r}-\bm{O}_{\xi})\bigr]\,
\rho_{\xi}(\bm{r}',t)\,
\mathrm{d}\lambda\,\mathrm{d}^3\bm{r}'\,,
\end{align}
where $Q_{\xi}=\int\rho_\xi(\bm{r},t)\,\mathrm{d}^3\bm{r}$ is the total charge of
molecule $\xi$. This formalism is particularly relevant for electrically neutral
molecules ($Q_{\xi}=0$).

From the above definitions and Eq.~\eqref{eq:constEQ_Mag}, it follows that the
magnetization field can be written in SI units as
\begin{equation}
\label{eq:M_general}
\bm{M}_{\xi}(\bm{r},t)
=
\int
\frac{\mu_0}{4\pi}\,
\frac{\nabla_{\bm{r}'}\times\bm{J}_{\xi}(\bm{r}',t)
-\nabla_{\bm{r}'}\times\bm{K}_{\xi}(\bm{r}',t)}
{\|\bm{r}-\bm{r}'\|}
\,\mathrm{d}^3\bm{r}'\,,
\end{equation}
where
\begin{equation}
\label{eq:K_def}
\bm{K}_{\xi}(\bm{r},t)
=
\int
\bm{\mathcal{G}}(\bm{r},\bm{r}';\bm{O}_{\xi})\,
\partial_t \rho_{\xi}(\bm{r}',t)\,
\mathrm{d}^3\bm{r}'\,.
\end{equation}
In general, the polarization and magnetization fields are origin dependent. This
could raise concerns regarding the applicability of such a representation to the
characterization of charge and current density distributions in matter.
However, this origin dependence does not affect physical couplings.

For instance, for an external electrostatic potential $\phi(\bm r)$, one has
\begin{equation}
\int \rho_{\xi}(\bm r)\,\phi(\bm r)\,\mathrm{d}^3\bm r
=
-\int \bigl[\nabla\cdot \bm P_{\xi}(\bm r;\bm O_\xi)\bigr]\,
\phi(\bm r)\,\mathrm{d}^3\bm r
=
-\int \bm P_{\xi}(\bm r;\bm O_\xi)\cdot \bm E(\bm r)\,\mathrm{d}^3\bm r,
\end{equation}
where $\bm E(\bm r)=-\nabla\phi(\bm r)$, and surface terms are assumed to vanish
(for localized charge distributions or appropriate boundary conditions).
The left-hand side is manifestly independent of the choice of $\bm O_\xi$;
therefore, the contracted form of the interaction is origin independent.
Analogous considerations apply to magnetostatic and fully electromagnetic
couplings.

We now show how the nonrelativistic QED Hamiltonian in the Coulomb gauge,
formulated in the minimal-coupling representation, can be mapped to an
equivalent Hamiltonian formulation in the multipolar coupling.

We begin with the minimal-coupling Lagrangian for a system of $N_{\rm mol}$
molecules labelled by $\xi$, each associated with a pole of reduction $\bm O_\xi$.
Molecule $\xi$ consists of $N_\xi$ nonrelativistic charged particles labelled by
$A_\xi=1,\dots,N_\xi$, interacting with the electromagnetic field,
\begin{align}
\label{eq:L_minimal}
\mathcal{L}_{\mathrm{min}}
=&
\sum_{\xi=1}^{N_{\rm mol}}\sum_{A_\xi=1}^{N_{\xi}}
\left[
\frac{1}{2} m_{A_\xi}\dot{\bm{r}}_{A_\xi}^{\,2}
+ q_{A_\xi}\dot{\bm{r}}_{A_\xi}\cdot \bm{A}(\bm{r}_{A_\xi},t)
- q_{A_\xi}\phi(\bm{r}_{A_\xi},t)
\right]
\nonumber\\
&\quad+
\frac{\varepsilon_0}{2}
\int\left[\|\bm{E}(\bm{r},t)\|^2-c^2\|\bm{B}(\bm{r},t)\|^2\right]\mathrm{d}^3\bm{r}\,.
\end{align}
Here $\phi(\bm{r},t)$ and $\bm{A}(\bm{r},t)$ are the scalar and vector potentials.
In the Coulomb gauge, $\nabla\cdot\bm{A}(\bm{r},t)=0$, the scalar potential
satisfies Poisson's equation
$\nabla^2\phi(\bm{r},t)=-\rho(\bm{r},t)/\varepsilon_0$, and the electric and
magnetic fields decompose as
\begin{align*}
E_{i}(\bm{r},t)
&=\sum_{j=1}^3\int
\left[\delta_{ij}^{\perp}(\bm{r}-\bm{r}')
+(\delta_{ij}-\delta_{ij}^{\perp}(\bm{r}-\bm{r}'))\right]E_j(\bm{r}',t)\,
\mathrm{d}^3\bm{r}'
\nonumber\\
&=E_{\perp,i}(\bm{r},t)+E_{\parallel,i}(\bm{r},t),\\
\bm{E}_{\parallel}(\bm{r},t)&=-\nabla\phi(\bm{r},t),
\qquad
\bm{E}_{\perp}(\bm{r},t)=-\partial_t\bm{A}(\bm{r},t),\\
\bm{B}(\bm{r},t)&=\nabla\times\bm{A}(\bm{r},t)\,.
\end{align*}

Using the decomposition of the current in terms of polarization and magnetization
fields, the same equations of motion can be obtained from the multipolar (PZW)
Lagrangian
\begin{align}
\label{eq:L_multipolar}
&\mathcal{L}_{\mathrm{PZW}}
=
\sum_{\xi=1}^{N_{\rm mol}}
\biggl[
\sum_{A_\xi=1}^{N_{\xi}}
\frac{1}{2} m_{A_\xi}\dot{\bm{r}}_{A_\xi}^{\,2}
+\sum_{B_{\xi} < A_{\xi}} V_{\rm Coul}(\bm{r}_{A_\xi},\bm{r}_{B_\xi})
+\sum_{\xi'< \xi}\sum_{B_{\xi '}=1}^{N_{\xi '}}
V_{\rm Coul}(\bm{r}_{A_{\xi}},\bm{r}_{B_{\xi'}})
\biggr]
\nonumber\\
&\quad+
\frac{\varepsilon_0}{2}
\int\left(\|\bm{E}(\bm{r},t)\|^2-c^2\|\bm{B}(\bm{r},t)\|^2\right)\mathrm{d}^3\bm{r}
+\int
\left[
\bm{P}(\bm{r},t)\cdot \bm{E}(\bm{r},t)
+
\bm{M}(\bm{r},t)\cdot \bm{B}(\bm{r},t)
\right]\mathrm{d}^3\bm{r}\,.
\end{align}

The two Lagrangians~\eqref{eq:L_minimal} and~\eqref{eq:L_multipolar} are related
by a total time derivative (the G\"oppert--Mayer term),
\begin{equation}
\label{eq:GM_term}
\mathcal{L}_{\mathrm{PZW}}
=
\mathcal{L}_{\mathrm{min}}
+
\frac{\mathrm{d}}{\mathrm{d}t}
\sum_{\xi=1}^{N_{\rm mol}}
\int
\bm{P}_{\xi}(\bm{r},t;\bm{O}_{\xi})\cdot \bm{A}(\bm{r},t)\,
\mathrm{d}^3\bm{r}\,.
\end{equation}

Passing to the Hamiltonian formulation requires defining the conjugate momenta of
the matter and field degrees of freedom, which read, respectively,
\begin{align}
p_{A_{\xi},i}
&=
\dfrac{\partial \mathcal{L}_{\rm min}}{\partial \dot{r}_{A_{\xi},i}}
=
m_{A_{\xi}}\dot{r}_{A_{\xi},i}
+
q_{A_{\xi}}A_{i}(\bm{r}_{A_{\xi}},t),
\\
\Pi_i(\bm{r})
&=
\dfrac{\partial \mathcal{L}_{\rm min}}{\partial \dot{A}_{i}(\bm{r})}
=
\varepsilon_{0}\dot{A}_{i}(\bm{r})
=
-\varepsilon_0 E_{\perp,i}(\bm{r})\,.
\end{align}
Applying the Legendre transform to the minimal-coupling Lagrangian, the
minimal-coupling Hamiltonian reads
\begin{align}
\label{eq:H_minimal}
H_{\mathrm{min}}
&=
\sum_{\xi=1}^{N_{\rm mol}}\sum_{A_\xi=1}^{N_{\xi}}
\frac{1}{2m_{A_\xi}}
\left\|
\bm{p}_{A_\xi}
-
q_{A_\xi}\bm{A}(\bm{r}_{A_\xi})
\right\|^2
+
\sum_{\xi=1}^{N_{\rm mol}}\sum_{A_\xi=1}^{N_{\xi}}
q_{A_\xi}\phi(\bm{r}_{A_\xi})
\nonumber\\
&\quad+
\frac{1}{2}\int
\left(
\frac{\|\bm{\Pi}(\bm{r})\|^2}{\varepsilon_0}
+ \varepsilon_0 c^2 \|\bm{B}(\bm{r})\|^2
\right)\mathrm{d}^3\bm{r}\,.
\end{align}

The canonical conjugate momenta derived from the Lagrangian in the multipolar
coupling read
\begin{align}
p_{A_{\xi},i}
&=
\dfrac{\partial \mathcal{L}_{\rm PZW}}{\partial \dot{r}_{A_{\xi},i}}
=
m_{A_{\xi}}\dot{r}_{A_{\xi},i}
-
\sum_{j,k}\int \epsilon_{ijk}\, n_{A_{\xi},j}(\bm{r})\, B_{k}(\bm{r})\,
\mathrm{d}^3\bm{r},
\\
\Pi_i(\bm{r})
&=
\dfrac{\partial \mathcal{L}_{\rm PZW}}{\partial \dot{A}_{i}(\bm{r})}
=
\varepsilon_{0}\dot{A}_{i}(\bm{r})
-\bm{P}^{\perp}_{i}(\bm{r})
=
-\bm{D}^{\perp}_{i}(\bm{r})\,,
\end{align}
where $\epsilon_{ijk}$ is the Levi-Civita symbol and $\bm{D}(\bm{r})$ is the
displacement field. Here $\bm{n}_{A_{\xi}}$ is a vector field of the form
\begin{equation}
\bm{n}_{A_{\xi}}(\bm{r})
=
-q_{A_{\xi}} (\bm{r}-\bm{O}_{\xi})
\int_{0}^1 \lambda\,
\delta\!\bigl(\bm{r}-\bm{O}_{\xi}-\lambda(\bm{r}_{A_{\xi}}-\bm{O}_{\xi})\bigr)\,
\mathrm{d}\lambda\,.
\end{equation}
By performing the Legendre transform of the Lagrangian $\mathcal{L}_{\rm PZW}$,
the multipolar Hamiltonian is given by
\begin{align}
\label{eq:H_PZW}
H_{\mathrm{PZW}}
&=
\sum_{\xi=1}^{N_{\rm mol}}\sum_{A_\xi=1}^{N_{\xi}}
\frac{\|\bm{p}_{A_\xi}\|^{\,2}}{2m_{A_\xi}}
+
\frac{1}{2\varepsilon_0}
\int
\left\|
\bm{\Pi}(\bm{r}) + \bm{P}^{\perp}(\bm{r})
\right\|^2
\mathrm{d}^3\bm{r}
\nonumber\\
&\quad+
\frac{1}{2\mu_0}
\int
\|\bm{B}(\bm{r})\|^2\,
\mathrm{d}^3\bm{r}
-
\int
\bm{M}(\bm{r},t)\cdot \bm{B}(\bm{r},t)\,
\mathrm{d}^3\bm{r}
\nonumber\\
&\quad+
\frac{1}{2}\sum_{i,j}\iint
B_{i}(\bm{r})\, O_{ij}(\bm{r},\bm{r}')\, B_{j}(\bm{r}')\,
\mathrm{d}^3\bm{r}\,\mathrm{d}^3 \bm{r}' \,,
\end{align}
where $O_{ij}(\bm{r},\bm{r}')$ is the diamagnetic interaction kernel,
\begin{equation}
O_{ij}(\bm{r},\bm{r}')
=
\sum_{\xi=1}^{N_{\rm mol}}\sum_{A_{\xi}=1}^{N_{\xi}}
\frac{1}{m_{A_{\xi}}}
\sum_{k,l,m}
\epsilon_{ikl}\epsilon_{jml}\,
n_{A_{\xi},k}(\bm{r})\,n_{A_{\xi},m}(\bm{r}')\,.
\end{equation}

It is important to note that the PZW Hamiltonian contains a self-energy term for
the polarization field,
$\tfrac{1}{2\varepsilon_0}\int \|\bm{P}^{\perp}(\bm{r},t)\|^2\,\mathrm{d}^3\bm{r}$,
which is particularly important because it contributes to making the Hamiltonian
bounded from below. Moreover, if the overlap of charge densities can be
neglected, it can be shown that the transverse and longitudinal components of
the direct interaction between polarization fields of two different molecules
mutually cancel, i.e.,
$\int \bm{P}^{\perp}_{\xi}(\bm{r})\cdot  \bm{P}^{\perp}_{\xi'}(\bm{r})
=-\int \bm{P}^{\parallel}_{\xi}(\bm{r})\cdot  \bm{P}^{\parallel}_{\xi'}(\bm{r})$.
It follows that the matter fields representing the molecular charge
distributions are coupled only to the electromagnetic field. For this reason,
the multipolar coupling framework is particularly suited to investigating
intermolecular interactions and, more generally, the QED of bounded systems of
charges.

The two Hamiltonians are related by a canonical transformation which, in the
quantum theory, is implemented by the unitary operator
\begin{equation}
\label{eq:U_GM}
\hat{U}_{\mathrm{GM}}(t)
=
\exp\!\left[
\frac{i}{\hbar}
\int
\hat{\bm{P}}(\bm{r},t)\cdot \hat{\bm{A}}(\bm{r},t)\,
\mathrm{d}^3\bm{r}
\right]\,.
\end{equation}
Here the two operators are evaluated at a fixed time $t$.

It should be noted that the length gauge, widely used in polaritonic chemistry
and cavity quantum electrodynamics, corresponds to the multipolar
representation truncated at electric-dipole order in the long-wavelength
approximation.

Accordingly, operators transform as
\[
\hat{O}_{\mathrm{PZW}}
=
\hat{U}_{\mathrm{GM}}\,
\hat{O}_{\mathrm{min}}\,
\hat{U}_{\mathrm{GM}}^{\dagger},
\]
and, in particular, the canonical field momentum transforms as
\[
\hat{\bm{\Pi}}^{(\mathrm{PZW})}(\bm{r})
=
\hat{U}_{\mathrm{GM}}\,
\hat{\bm{\Pi}}^{(\mathrm{min})}(\bm{r})\,
\hat{U}_{\mathrm{GM}}^{\dagger}
=
\hat{\bm{\Pi}}^{(\mathrm{min})}(\bm{r})
+
\hat{\bm{P}}^{\perp}(\bm{r})\,.
\]

The transverse canonical commutation relations of the electromagnetic field
\begin{equation}
\label{eq:CCR}
\left[
\hat{A}_i(\bm{r}),
\hat{\Pi}_j(\bm{r}')
\right]
=
i\hbar\,
\delta^{\perp}_{ij}(\bm{r}-\bm{r}')\,,
\end{equation}
where $\delta^{\perp}_{ij}(\bm{r}-\bm{r}')$ is the dyadic projector operator, are
preserved under the transformation, since electromagnetic-field operators commute
with matter operators. For further details on multipolar coupling, we refer the
reader to the literature~\cite{Craig1994,Salam2009,woolley2022foundations}.

\bibliography{merged_unique}
\end{document}